%% file: Nearly_Optimal_Sparse_Group_Testing.tex
\documentclass[11pt]{article}

\usepackage{amsthm}
\usepackage{amsmath,amsfonts,amssymb,latexsym,cite}
\usepackage{soul}

\usepackage{enumerate}
\usepackage{hyperref} 
\usepackage{graphicx} 
\graphicspath{{./figures/}} 
\usepackage{subcaption}
\usepackage{color}
\usepackage{epsfig}
\usepackage{graphicx}
\usepackage{multirow}
\usepackage{latexsym}
\usepackage{amsmath}
\usepackage{amssymb}
\usepackage{fullpage}
\usepackage{tikz}
\usepackage[authormarkup=none,final]{changes}
\usepackage{rotating}
\usepackage{float}

\definechangesauthor[name={GV},color=red]{gv}
\definechangesauthor[name={Elena},color=blue]{eg}
\definechangesauthor[name={Samson},color=orange]{sz}
\definechangesauthor[name={Sid},color=magenta]{sj}

\setaddedmarkup{#1~}

\newif\ifnotes\notestrue 
\ifnotes
\newcommand{\elena}[1]{\textcolor{red}{{\bf (Elena:} {#1}{\bf ) }} \marginpar{\tiny\bf
             \begin{minipage}[t]{0.5in}
               \raggedright E:
                \end{minipage}}}
\newcommand{\samson}[1]{\textcolor{green}{{\bf (Samson:} {#1}{\bf ) }} \marginpar{\tiny\bf
             \begin{minipage}[t]{0.5in}
               \raggedright S:
            \end{minipage}}}
\newcommand{\gv}[1]{\textcolor{purple}{{\bf (GV:} {#1}{\bf ) }} \marginpar{\tiny\bf
             \begin{minipage}[t]{0.5in}
               \raggedright G:
            \end{minipage}}}  
\newcommand{\sid}[1]{\textcolor{blue}{{\bf (Sid:} {#1}{\bf ) }} \marginpar{\tiny\bf
             \begin{minipage}[t]{0.5in}
               \raggedright J:
            \end{minipage}}} 
\else
\newcommand{\elena}[1]{}
\newcommand{\samson}[1]{}
\newcommand{\gv}[1]{}
\newcommand{\sid}[1]{}
\fi

\hypersetup{
     colorlinks   = true,
     citecolor    = blue,
		 linkcolor		= red
}

\newtheorem{theorem}{Theorem}[section]

\newtheorem{fact}[theorem]{Fact}

\theoremstyle{remark}
\newtheorem{remark}[theorem]{Remark}

\newenvironment{proofof}[1]{\begin{trivlist} \item {\bf Proof
#1:~~}}
  {\qed\end{trivlist}}

\begin{document}
	\title{Nearly Optimal Sparse Group Testing\footnote{A preliminary version \cite{GandikotaGJZ16} of this paper appeared in the Proceedings of the 54th Annual Allerton Conference on Communication, Control, and Computing (Allerton 2016).
}}
\author{
Venkata Gandikota\thanks{Department of Computer Science, Johns Hopkins University, Baltimore, MD. 
Email: {\tt gv@jhu.edu}.}
\and
Elena Grigorescu\thanks{Department of Computer Science, Purdue University, West Lafayette, IN. 
Email: {\tt elena-g@purdue.edu}.}
\and
Sidharth Jaggi\thanks{Department of Information Engineering, The Chinese University of Hong Kong, Shatin, N.T., Hong Kong. 
Email: {\tt jaggi@ie.cuhk.edu.hk}.}
\and
Samson Zhou\thanks{School of Informatics, Computing, and Engineering, Indiana University, Bloomington, IN. 
This work was done in part while at the Department of Computer Science, Purdue University, West Lafayette, IN. 
Email: {\tt samsonzhou@gmail.com}.}
}

\maketitle
\begin{abstract}
	\input{abstract}
\end{abstract}
\input{intro}
\input{background}
\input{model}
\input{proof-outline}
\input{results}
\input{column-information}
\input{column-construction}
\input{column-explicit}
\input{row-information}
\input{row-construction}
\input{row-explicit}
\input{noise}
\input{futurework}
\input{acknowledgements}
\bibliographystyle{plainyr}
\bibliography{./gem}

\end{document}

%% file: abstract.tex
Group testing is the process of pooling arbitrary subsets from a set of $n$ items so as to identify, with a minimal number of tests, a ``small'' subset of $d$ defective items. 
In ``classical'' non-adaptive group testing, it is known that when $d$ is substantially smaller than $n$, $\Theta(d\log(n))$ tests are both information-theoretically necessary and sufficient to guarantee recovery with high probability. 
Group testing schemes in the literature meeting this bound require most items to be tested $\Omega(\log(n))$ times, and most tests to incorporate $\Omega(n/d)$ items.

Motivated by physical considerations, we study group testing models in which the testing procedure is constrained to be ``sparse''. Specifically, we consider (separately) scenarios in which (a) items are finitely divisible and hence may participate in at most $\gamma \in o(\log(n))$ tests; or (b) tests are size-constrained to pool no more than $\rho \in o(n/d)$items per test. 
For both scenarios we provide information-theoretic lower bounds on the number of tests required to guarantee high probability recovery. 
In particular, one of our main results shows that $\gamma$-finite divisibility of items forces {\it any} non-adaptive group testing algorithm with probability of recovery error at most $\epsilon$ to perform at least $\gamma d(n/d)^{(1-5\epsilon)/\gamma}$ tests. 
Analogously, for $\rho$-sized constrained tests, we show an information-theoretic lower bound of $\Omega(n/\rho)$ tests -- hence in both settings the number of tests required grow dramatically (relative to the classical setting) as a function of $n$. 
In both scenarios we provide both randomized constructions (under both $\epsilon$-error and zero-error reconstruction guarantees) and explicit constructions of designs with computationally efficient reconstruction algorithms that require a number of tests that are optimal up to constant or small polynomial factors in some regimes of $n, d, \gamma \text{ and } \rho$. 
The randomized design/reconstruction algorithm in the $\rho$-sized test scenario is {\it universal} -- independent of the value of $d$, as long as $\rho \in o(n/d)$. We also investigate the effect of unreliability/noise in test outcomes.

%% file: intro.tex
\section{Introduction} \label{sec:introduction}
The problem of group testing deals with identifying a relatively small number of ``defective'' items among a large population via non-linear ``grouped'' tests. 
The model was introduced by Dorfman \cite{Dorfman43} in 1943, motivated by the task of identifying syphilitic individuals among military inductees during World War II. 
Individual blood tests for syphilis were expensive, so the idea was to pool and test multiple blood samples simultaneously. 
It was desirable to minimize the number of tests, while correctly identifying the disease status of every individual. 

This paper studies group testing with two potential types of constraints on the group-testing procedure. 
First, we consider a model wherein each item can be tested a limited number $\gamma$ of times ({\it e.g.} due to a limited amount of blood that can be taken from an individual). 
Second, we consider a model wherein each test can be a pool of at most a certain number $\rho$ of items ({\it e.g.} equipment limitations may impose a maximum on the number of objects that can be simultaneously tested). 

Our primary technical contribution in this work is to demonstrate that even relatively mild constraints on $\gamma$ or $\rho$ can dramatically change the nature of the group-testing problem. 
In contrast to the classical (unconstrained) group-testing literature, where $\Theta(d\log(n))$ tests are necessary and sufficient for high probability recovery, in our ($\gamma$- or $\rho$- constrained) setting the necessary number of tests required may be as much as an {\it exponential} factor (in $n$) larger (polynomial in $n$ rather than logarithmic in $n$)! 
We also present a suite of algorithms (in a variety of problem settings) that meet these information-theoretic bounds up to constant or small polynomial (in $d$ or $n$) factors. 
The randomized design/reconstruction algorithm in the $\rho$-sized test scenario may also be made {\it universal} -- independent of the value of $d$, as long as $\rho \in o(n/d)$.

%% file: background.tex
\subsection{Related work} \label{sec:background} 
While there is significant literature on multiple alternative models of group testing (see for instance\cite{Bonis14, CaiJBJ13, ChanCBJS13, CheraghchiKMS12, CheraghchiHKV11, ESM11, LiCHKJ14, ScarlettC16,lee2016saffron, IndykNR10}),
the focus of this work is primarily on non-adaptive group testing (NAGT), under $\epsilon$-error and zero-error reconstruction guarantees. 
We thus restrict the  discussion of prior work to the literature on lower bounds and algorithms (both deterministic and randomized) for $\epsilon$-error and zero-error non-adaptive group testing.

Du and Hwang \cite{DuH00} show that ${\cal O}(d^2\log n)$ tests suffice for zero-error group testing, while Porat and Rothschild \cite{PoratR08} provide an explicit NAGT algorithm with ${\cal O}(d^2\log n)$ tests, almost matching the best known lower bound of $\Omega\left(\min\{ n,d^2\log n / \log d \} \right)$ \cite{ErdosFF85}. 

The lower bound of $(1-\epsilon)d\log(n/d)$ for  $\epsilon$-error group testing, \cite{ChanJSA14} is met (up to constant factors) by \cite{Sebo85, ScarlettC16, AldridgeJS16}.

In all the works mentioned above there are no \emph{a priori}constraints on the group tests themselves.
In classical group testing algorithms that meet (up to constant factors) the information-theoretic lower bound of $\Omega(d\log(n))$ tests for  $\epsilon$-error reconstruction, each item is tested $\Omega(\log(n/d))$ times. 

\subsubsection{Most relevant papers}\label{subsubsec:most-relevant}
In this section we discuss in somewhat more detail prior work that is potentially the closest to this paper.
\begin{itemize}
\item
While an earlier conference version of this paper~\cite{GandikotaGJZ16} also had results on zero-error designs in the $\gamma$-divisible items and $\rho$-sized test settings, subsequent work in~\cite{InanKO17} subsequently significantly improved on those results with nearly matching achievabilities and lower bounds in the zero-error setting. We hence reference readers interested in zero-error constrained combinatorial designs to~\cite{InanKO17}. We note in passing that while a direct comparison between the results in~\cite{InanKO17} and this work is unfair due to the difference in error criteria (vanishing error versus zero-error), the number of tests required for some designs in our work are {\it significantly} smaller than corresponding results in~\cite{InanKO17} (for instance an {\it upper} bound of ${\cal O}\left (\gamma d(n/\epsilon)^{1/\gamma}\right)$ tests sufficing to ensure probability of error of at most $\epsilon$ in Theorem~\ref{thm:col-ub-random} of this work, versus a {\it lower} bound of about $\Omega\left ((nd^2)^{d/\gamma}\right)$ in~\cite{InanKO17}). This significant penalty that one pays when one demands zero-error performance (instead of being satisfied with vanishing error performance) is analogous the difference between information-theoretic and coding-theoretic results in general, and to corresponding known results in classical non-adaptive group testing, where $\Theta(d\log(n))$ tests are sufficient (and necessary for vanishing-error performance, versus $\Omega\left(\min\{ n,d^2\log n / \log d \} \right)$ \cite{ErdosFF85} being necessary for zero-error performance.
\item
Perhaps the first work to consider group-testing with constraints on column or row weights of the design matrix was~\cite{macula96}, wherein a very elegant construction  of {\it both} row and column constraints was provided under a zero-error of reconstruction criteria. However, the number of tests required is significantly greater than those required in this work. Specifically, viewed as a $\gamma$-constrained design, the number of tests required in~\cite{macula96} scales as ${\cal O}\left (n^{d/\gamma^{1/d}}\right )$, which is significantly larger than the ${\cal O}\left (\gamma d n^{1/\gamma}\right )$ tests that suffice in the randomized design (Theorem~\ref{thm:col-ub-random}) in this work, or the designs in~\cite{macula96}. Alternatively, viewed as a $\rho$-constrained design, the number of tests required in~\cite{macula96} scales as ${\cal O}\left (\frac{n}{\rho} \left (\frac{\log(n)}{\log(n\rho)}\right )^d \right )$, which is significantly larger than the ${\cal O}\left (\frac{n}{\rho}  \right )$ tests that suffice in the randomized design (Theorem~\ref{thm:row-ub-random}) in this work.
\item
The title of  \cite{AldridgeBJ} is ``Improved group testing rates with constant column weight designs'', and as such it may be tempting to consider this work as related to the column-constrained group-testing problem. While~\cite{AldridgeBJ} is indeed very nice work for other reasons outlined below, in actuality is not directly relevant to the problem at hand since the value of $\gamma$ chosen is $\Theta(\log(n))$. Hence the word ``constant'' does not truly mean a constant independent of $n$ (as can be the case in our setting when $\gamma \in \Theta(1)$.) The reason for this choice of $\gamma$ by the authors of \cite{AldridgeBJ} is that this allows them to improve over prior work by a constant factor the number of tests required for successful reconstruction, from $c_1d\log(n/d)$ to $c_2d\log(n/d)$, where $c_2$ is a somewhat smaller constant than $c_1$ (both constants depending on the limiting behaviour $\log(d)/\log(n)$). However, as we now note in our discussion in Section~\ref{sec:results}, $\gamma \in \Theta(\log(n))$ is no longer ``sparse'' -- if $\gamma$ is {\it substantially} smaller ($\gamma \in o(\log(n))$ then indeed the number of tests required may increase by a factor that is as much as exponentially larger in $n$).
\item 
Certain works such as~\cite{mezard2008group} focused on graph-based test-matrix ensembles with fixed row/column weights, and used insights from statistical physics (in particular via the ``replica symmetric cavity method'') to analyze their performance.~\footnote{Another work in this vein was~\cite{Wadayama17}, which was later discovered to have an error, and was withdrawn~\cite{wadayama2018comments}.} 
However, full proofs of correctness in these works have been omitted.
\end{itemize}

%% file: model.tex
\section{Model} \label{sec:model}
Let $\mathcal{S}$ be the set of $n$ elements, and let set $\mathcal{D}\subset \mathcal{S}$ with $|\mathcal D|=d$ consist of the {\em defective} items. 
The elements in $\mathcal S /\mathcal D$ are called {\em non-defective}.
Here $d$ is considered to be ``small'' with respect to $n$ -- it could be as small as a constant, or as large as $n^{\alpha}$ for some constant $\alpha$ strictly less than $1$.\footnote{It has recently been shown~\cite{agarwal2018novel,aldridge2018individual,wadayama2018comments} that for classical (unconstrained) group testing, in the regime $d \in \Theta(n)$ individual testing is optimal when one desires vanishing probability of reconstruction error -- i.e., there is no ``group-testing gain''. The regime where $d$ is {\it almost} linear in $n$, such as $d \in \Theta(n/\log(n))$ is still open -- the best upper and lower bounds differ by super-constant factors~\cite{ScarlettC16}. Hence in the constrained setting considered in this work we also restrict ourselves to the regime where $d$ scales polynomially slower than $n$.}

We wish to identify the defective items through a series of {\it group tests}, which takes as input a subset (group) of the $n$ items, and output whether or not there exists at least one defective item in the subset (group).\footnote{In classical group testing, the true value of $d$ (or a (good) upper bound on it) is typically known \emph{a priori}. 
Our algorithms in general only require knowledge of a (good) upper bound on $d$ -- they are designed under the assumption that the number of defectives is {\it at most} $d$, and will still have low probability of reconstruction error if the true number of defectives is any smaller value. 
Indeed, in some scenarios (for instance see Remark~\ref{rem:universal}) our designs and reconstruction algorithms are {\it universal}, in the sense that they do not need to know the value of $d$ at all, as long as one is guaranteed that $d \in o(n/\rho)$.}
The goal of non-adaptive group testing is to correctly identify the exact set of defective items with a minimum number of non-adaptive group tests.

Group testing may be \emph{adaptive} (the set of items to be tested in a group may be a function of prior test outcomes) or \emph{non-adaptive} (all group tests have to be chosen independently of prior test outcomes). 
Aside from naturalness of the non-adaptive group-testing problem viewed as a nonlinear estimation problem, the advantage of non-adaptive group tests over adaptive group tests is that they allow for parallel testing, and can use off-the-shelf hardware. 
Another reason to prefer non-adaptive group-testing over adaptive group-testing is its application as a module in other algorithms, for instance in streaming algorithms~\cite{cormode2005s}. 
We thus focus on non-adaptive group testing in this paper.

Formally, we represent the set $\mathcal{S}$ by the weight-$d$ binary \emph{input vector} $X\in\{0,1\}^n$, which is the indicator vector of $\mathcal S$. For a vector $v\in \{0,1\}^n$ we let $Supp(v)$ denote the set of non-zero indices in $v$.

The sequence of $T$ non-adaptive tests is represented by the rows $M_{i} \in \{0,1\}^n$ of a {\em test matrix} $M\in \{0, 1\}^{T\times n}$, and the outcomes $Y_{i}\in\{0,1\}$ are computed as $Y_{i}=\bigvee_{i\in Supp(M_{i})} X_i$, for all $i\in [T]$. 
Therefore, the result of a test is positive if and only if the test contains at least one defective item. We note that the test matrix could be chosen {\it deterministically}, or may be {\it randomized}.

Given the tests and their outcomes, the decoding algorithm outputs an \emph{estimate vector} $\widehat{X}\in\{0, 1\}^n$,  representing an estimate of the indicator vector of the defectives. 

{We are interested in tests that fail with small probability. 
In particular, we focus on {\em $\epsilon$-group testing} and  design tests for which the probability of error is bounded by some $\epsilon>0$\footnote{Note that in our proofs $\epsilon$ does double duty as both the probability of error, and as a ``generic sufficiently small positive quantity'' whenever some slack is needed. 
This allows us to reduce the number of parameters to keep track of, and eases exposition.  
}, namely  $$P_{error}={\mathbf{Pr}}\big[\widehat{X}\neq X\big]<\epsilon.$$ 
The probability in the above definition is taken over the randomness of the set of defectives, and over the randomness of the test matrix (if the tests are randomized). 
Here we assume that the set ${\mathcal D}$ of defectives is chosen uniformly at random among all sets of size $d$, {\it except} in Theorem~\ref{thm:column-noise}/Section~\ref{subsec:column-noise}, where for ease of analysis we consider instead the closely related model wherein items are defective i.i.d. Bernoulli($d/n$), leading to an expected group size of $d$. 
\footnote{Another slightly different distributions over $\mathcal{D}$ that might be considered to have a uniform distribution over all $\sum_{i=0}^d\binom{n}{i}$ subsets of size {\it at most} $d$ (rather than {\it exactly} $d$, as we do in our model). 
It turns out that these model perturbations do not substantially change results in the classical group-testing literature, and hence in this work we focus on just the model wherein each set of $d$ items may equal ${\mathcal D}$ with probability $1/\binom{n}{d}$, except in Theorem~\ref{thm:column-noise}/Section~\ref{subsec:column-noise}, where items are defective i.i.d. Bernoulli($d/n$).}}

Some authors  \cite{BaldassiniJA13, CheraghchiHKV11}  consider ``noisy'' tests, in which test outcomes are misreported with some small probability. 
In this setting, the output vector $\widehat{Y}\in \{0,1\}^T$ is obtained as $\widehat{Y}=Y \oplus Z$, where $Z\in\{0,1\}^T$ is a noise vector produced by a binary-symmetric channel with crossover probability $\sigma$. 
In such models, an $\epsilon$-error reconstruction guarantee is desired again, with $P_{error}={\mathbf{Pr}}\big[\widehat{X}\neq X\big]<\epsilon$, where the probability is taken over the randomness of $X$, possibly that of $M$, and that of {\em  the noise process that converts $Y$ to a noisy vector $\widehat{Y}$}. 
We briefly consider such models in Section~\ref{sec:noise}.

We define the \emph{$\gamma$-divisible group-testing} model as one in which each item can be tested at most $\gamma$ times, and so each column of $M$ contains at most $\gamma$ many 1's. 
Similarly, we define the \emph{$\rho$-sized group-testing} model as one in which each test can include at most $\rho$ items, and so each row of $M$ contains at most $\rho$ many 1's.

In what follows, all logarithms are base $2$. 
The function $H(X)$ denotes the entropy of the (vector valued) random variable $X$, $H(p)$ denotes the binary entropy function, and $I(X;Y)$ the mutual information between $X$ and $Y$. 

We first reprise the well-known {\it Stirling approximation} for the factorial function.
\begin{fact}[{\bf Stirling's approximation~\cite{de1981asymptotic}}]\label{fact:stirling-approx}
The factorial function $n!$ can be bounded from above and below as $$\left (1 -{\cal O}\left (\frac{1}{n} \right )\right )\sqrt{2\pi n}\left (\frac{ne}{d} \right)^n \leq  n! \leq \left (1 +{\cal O}\left (\frac{1}{n} \right )\right )\sqrt{2\pi n}\left (\frac{ne}{d} \right)^n.$$
\end{fact}

We use the following bounds~\cite{cormen2009introduction} on the binomial coefficients which follow from Stirling's approximations
\begin{fact}\label{fact:stirling}({\bf Bounds on binomial coefficients}~\cite{cormen2009introduction})
For any integers $n>0$, and $d \leq n$, 
\[ \left( \frac{n}{d} \right)^{d} \leq \binom{n}{d} \leq \left( \frac{en}{d} \right)^{d}.\]
\end{fact}

%% file: proof-outline.tex
\section{Outline}
\label{sec:proof-outline}
Before the formal statements (in Section~\ref{sec:results}) and proofs (in subsequent sections) of our results, we first present in this section a summary of our techniques, with supplementary intuition.

For our information-theoretic lower bounds on $\epsilon$-error non-adaptive group testing (in Sections~\ref{sec:col:lb} and \ref{subsec:row-lb}), we start with the observation that for any group testing procedure to succeed, the entropy of the test outcome vector $Y$ must almost equal the entropy of the input vector $X$ (which has entropy $\log\binom{n}{d}$, which is approximately $d\log(n/d) + \mathcal{O}(d)$ for $d=o(n^{\alpha})$) for any positive constant $\alpha $ in $ [0,1)$. (Note that if $d$ is say a constant, or scales logarithmically in $n$, this may be interpreted as $\alpha$ being an arbitrarily small positive constant.). 
Indeed, in classical group testing, this is a design principle for the test matrices $M$, leading to designs such that the probability of test outcomes being either positive or negative are each close to $1/2$ (and hence the entropy of each individual test is close to $1$).\footnote{Note that this is not a sufficient condition to guarantee low-error reconstructability of $X$ from $Y$, merely a necessary one. 
For instance, consider a test matrix $M$ such that the first test $Y_1$ has entropy $1$ bit, and each of the remaining $d\log(n/d)$ rows are identical to this first row. 
So while the sum of the entropies of individual tests is large, the overall entropy of the test outcome vector is {\it  just} 1 bit. 
This is due to the {\it extreme} correlation across tests. 
So really, one needs to design a matrix $M$ which not only has high entropy per tests, but also high entropy for most collections of tests. 
Nonetheless, as a lower-bounding technique, bounding the entropy of individual tests often provides a reasonable first-order approximation, as indeed seems to be the case in this work.}
This design principle implies that each test should include about $g^*=(n/d)\ln(2)$ items, since then the probability of a negative test outcome 
can then be shown to be $\approx 1/2$.
This density of items per test (corresponding to the density of items per row of $M$) 
coupled with the desire to use only an information-theoretically optimal number of tests of about $\Theta(d\log(n/d))$ (hence restricting $M$ to have $\Theta(d\log(n/d))$ rows), induces the fact that each column of $M$ should have on average about $\Theta(\log(n/d))$ items.

But for ``sparse'' matrices, for instance when tests are size-constrained to $\rho = o(n/d)$, it may be impossible to meet this design principle. This implies a fundamental upper bound on the entropy that can be ``squeezed'' out of each test $Y_i$. 
Coupled with the need to squeeze a total of $d\log(n/d)$ bits of entropy out of the test vector $Y$, and ``standard'' information-theoretic techniques (such as Fano's inequality) relating entropic quantities to probability of error give us non-trivial lower bounds on the number of tests required in the $\rho$-size constrained model, as outlined in Section~\ref{subsec:row-lb}.

Similar techniques also work in Section~\ref{sec:col:lb} to provide lower bounds in the case when the testing procedure involves $\gamma$-divisible items -- this puts a fundamental upper bound on the number of $1$'s in any column of the testing matrix $M$. This implies a constraint on the {\it average} density of each row in $M$. 
However, more care is required in this model, since there may be a few test rows of $M$ with ``high'' weight. 
Our bounding technique therefore proceeds by choosing a threshold above which we consider a test to be ``heavy''. 
We then do a two-stage approximation to obtain an upper bound on the entropy of the test outcome vector $Y$, and the rest of the proof is similar to the one 
in Section~\ref{subsec:row-lb}.

As an explicit example of the type of results obtainable via these lower bounding techniques, we can show that to detect a single defective ($d=1$) out of $n$ items, with a constraint that each item may be tested at most twice ($\gamma = 2$), it must be the case that group testing procedure has at least about $\sqrt{n}$ tests. (Compared with $\log(n)$ tests, which would suffice in the unconstrained case.)
To gain further intuition on why such a lower bound might be tight, consider the following testing algorithm. 
\begin{figure}[!ht]
\centering
\begin{tikzpicture}[scale=1]
\draw[step=1cm] (0,0) grid (3,3);
\draw [fill=red] (2cm,1cm) rectangle (3cm,2cm);
\node at (0.5cm,0.5cm){\Huge 6};
\node at (1.5cm,0.5cm){\Huge 7};
\node at (2.5cm,0.5cm){\Huge 8};
\node at (0.5cm,1.5cm){\Huge 3};
\node at (1.5cm,1.5cm){\Huge 4};
\node at (2.5cm,1.5cm){\Huge 5};
\node at (0.5cm,2.5cm){\Huge 0};
\node at (1.5cm,2.5cm){\Huge 1};
\node at (2.5cm,2.5cm){\Huge 2};
\draw (0.5cm,3.5cm) -- (0.5cm,3cm);
\draw (1.5cm,3.5cm) -- (1.5cm,3cm);
\draw (2.5cm,3.5cm) -- (2.5cm,3cm);
\draw [->](0.5cm,0cm) -- (0.5cm,-0.5cm);
\draw [->](1.5cm,0cm) -- (1.5cm,-0.5cm);
\draw [->](2.5cm,0cm) -- (2.5cm,-0.5cm);
\node at (0.3cm,3.5cm)[above]{Test 1};
\node at (1.5cm,3.5cm)[above]{Test 2};
\node at (2.7cm,3.5cm)[above]{Test 3};
\draw (-0.5cm,0.5cm) -- (0,0.5cm);
\draw (-0.5cm,1.5cm) -- (0,1.5cm);
\draw (-0.5cm,2.5cm) -- (0,2.5cm);
\draw [->](3cm,0.5cm) -- (3.5cm,0.5cm);
\draw [->](3cm,1.5cm) -- (3.5cm,1.5cm);
\draw [->](3cm,2.5cm) -- (3.5cm,2.5cm);
\node at (-0.5cm,0.5cm)[left]{Test 6};
\node at (-0.5cm,1.5cm)[left]{Test 5};
\node at (-0.5cm,2.5cm)[left]{Test 4};
\node at (0cm,-0.5cm)[below]{negative};
\node at (1.5cm,-0.5cm)[below]{negative};
\node at (3cm,-0.5cm)[below]{positive};
\node at (3.5cm,0.5cm)[right]{negative};
\node at (3.5cm,1.5cm)[right]{positive};
\node at (3.5cm,2.5cm)[right]{negative};
\end{tikzpicture}
{\caption{If $n=9,\gamma=2,d=1$, the above test uniquely determines that item $5$ is defective.}\label{fig:toy}}
\end{figure}
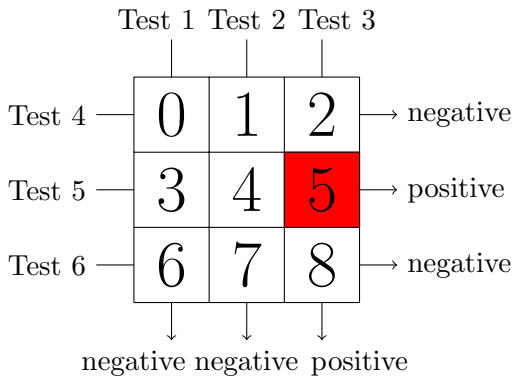
The $n$ items are arranged into a $\sqrt{n} \times \sqrt{n}$ grid, as in Figure~\ref{fig:toy}. 
The test matrix then comprises of $2\sqrt{n}$ rows, corresponding to the $\sqrt{n}$ sets of $\sqrt{n}$ items in each column of this grid, and the $\sqrt{n}$ sets of $\sqrt{n}$ items in each row of this grid. 
The unique defective item then must correspond to the item sitting at the intersection of the single column and the single row that return positive test outcomes.

Generalizing this explicit design to general item and test constrained settings takes more work. 
We provide explicit constructions that use the toy example ($d = 1, \gamma = 2$) and generalize to arbitrary $d$, and $\gamma$ or $\rho$, via a ``divide-and-conquer'' approach. 
Details are provided in Sections~\ref{sec:col:exp} and \ref{sec:row:exp}.

In Sections~\ref{sec:col:random} and~\ref{sec:row:random} we also provide randomized designs that draw intuition from the analysis of ``classical'' (unconstrained) group testing schemes. 
We analyze the probability that randomly chosen matrices chosen from suitable ensemble of matrices with either $\rho$-sparse rows or $\gamma$-sparse columns (for the two models considered) have a ``reasonable'' probability of success, by analyzing the probability that a non-defective item is ``masked'' by the set of $d$ defective items. 
These results are outlined in Sections~\ref{sec:col:random} and~\ref{sec:row:random}. We highlight a special feature of the result in Section~\ref{sec:row:random}, that it is universal, in the sense that the corresponding design/reconstruction algorithms are {\it independent} of the value of $d$ (as long as $\rho$ is in the range of interest, i.e., $\rho \in o(n/d)$.

Finally, in Section~\ref{sec:noise} we examine the effect of $\sigma$-noise (say BSC($\sigma$) noise for concreteness) in test outcomes $Y_i$ on the reconstructability of $X$ -- interestingly, while non-trivial achievability schemes exist in the $\rho$-test size constrained setting with $\sigma$-noisy test outcomes (for instance by repeating each test an appropriate number of times and taking the majority), in the $\gamma$-divisible item scenario any non-trivial amount of noise renders {\it any} group testing algorithm unable to reconstruct $X$ with a vanishing probability of error. 
This latter impossibility result stems from the fact that if the columns of $M$ are sufficiently sparse ($o(\log(n)$), then with non-trivial probability $(1-\sigma^{o(\log(n))})^n$, all information about the status of at least one item will be completely masked by the noise in the tests in which the item participates.

%% file: results.tex
\section{Results}\label{sec:results}
We now formally state our results, with proofs presented in subsequent sections.
\subsection{{$\gamma$-divisible items}} 
We begin with discussing our results for the scenario wherein each item may be tested at most $\gamma$ times.

\subsubsection{Parameter regime of interest for $\gamma$-divisible items}\label{subsec:col-para}
Some prefatory remarks that will be useful in the proofs are in order. 
It is well-known in the classical group-testing literature (see for instance~\cite{ChanJSA14,ScarlettC16}) that in unconstrained settings $T >(1-\epsilon)(d\log(n/d))$ tests are necessary to reconstruct all defectives with error probability at most $\epsilon$, hence the same is certainly true in the constrained setting. 
It is also known that if each item can be tested $\Theta(\log(n/d))$ times, then $\Theta(d\log(n))$ tests suffice for group-testing algorithms with vanishing error.
Hence the parameter regime of primary interest is when $\gamma \in o(\log(n/d)) = o(\log(n))$ (since $d \in o(n^{\alpha})$ for some positive constant $\alpha < 1$), and when $T \in \Omega(d\log(n/d))$. Note that this implies a fact that will be useful later in our proofs, that 
\begin{equation}
\frac{T}{\gamma d} \in \omega(1).\label{eq:Tgammad}
\end{equation}
To trade off between precision and readability, we specify explicit (upper and lower) bounds in the following theorems, but use instead Bachmann-Landau asymptotic notation to specify parameter regimes and computational complexity.

\subsubsection{Results for $\gamma$-divisible items}\label{subsec:col-results}
Perhaps the result with the most technically involved proof involves a lower bound (Theorem~\ref{thm:col-lb}) on the number of tests in any non-adaptive group-testing algorithm that tests each item at most $\gamma \in o(\log(n))$ times, and is allowed to make an error with probability at most $\epsilon$. This result sets the stage for the remainder of this work, by showing the stark price one pays for imposing constraints on test designs, with the number of tests shooting up from being logarithmic in $n$ in the unconstrained setting, to being polynomial in $n$ in the constrained setting. 
\begin{theorem}[Section~\ref{sec:col:lb}]
\label{thm:col-lb}
For any sufficiently large $n$, sufficiently small $\epsilon>0$, and $d \in o(n^\alpha)$ for some positive constant $\alpha < 1$, any non-adaptive group-testing algorithm that tests each item at most $\gamma \in o(\log n)$ times and has a probability of error of at most $\epsilon$ requires at least $\gamma d\left(\frac{n}{d}\right)^{\frac{1-5\epsilon}{\gamma}}$ tests.
\end{theorem}

\noindent To complement this lower bound, we also present two designs, both of which have affiliated reconstruction algorithms with computational complexities that are polynomial in problem parameters.

The first provides a randomized construction, requiring a number of tests that is larger than the lower bound by a factor that scales essentially as $\Theta(\left (\frac{d}{\epsilon}\right )^{1/\gamma})$ (neglecting lower-order dependencies on the probability of error $\epsilon$) -- see Remark~\ref{rem:col:ratio} for a discussion on the potential reasons for this gap. 
\begin{theorem}[Section~\ref{sec:col:random}]
\label{thm:col-ub-random}
For any sufficiently large $n$, sufficiently small $\epsilon >0$, and $d \in o(n^\alpha)$, for some positive constant $\alpha < 1$, there exists a randomized design testing each item at most $\gamma \in o(\log(n))$ times that uses at most $\left\lceil e\gamma d\left(\frac{n}{\epsilon}\right)^{\frac{1}{\gamma}}\right\rceil$ tests, and an affiliated reconstruction algorithm of computational complexity ${\cal O}\left ( \gamma d\left(\frac{n}{\epsilon}\right)^{1+1/\gamma}\right )$, that ensures a reconstruction error of at most $\epsilon$.
\end{theorem}

The second design is explicit (does not require randomness in designing the test matrix). 
On the one hand this design requires more tests than the one in Theorem~\ref{thm:col-ub-random} (by a factor of $\Theta\left(\frac{d}{\epsilon} \right)^{1-2/\gamma}$), but on the other hand for small $d$ (say constant $d$, or $d \in {\cal O}(\log n)$) has reconstruction complexity that can be {\it exponentially} smaller in $n$ than that of the algorithm in Theorem~\ref{thm:col-ub-random}. 
One point to highlight about Theorem~\ref{thm:col-ub-explicit} is that due to technical reasons, the construction only works when $d \in o\left (\sqrt{n} \right )$, rather than $d \in o(n)$ as in Theorem~\ref{thm:col-ub-random}.

\begin{theorem}[Section~\ref{sec:col:exp}]
\label{thm:col-ub-explicit}
For any sufficiently large $n$, sufficiently small $\epsilon >0$, and $d \in o(n^\alpha)$ for some positive constant $\alpha < 1/2$, there exists a deterministic design testing each item at most $\gamma \in o(\log(n))$ times that uses at most $\lceil\frac{d^2\gamma}{\epsilon}\rceil\left \lceil (\frac{n\epsilon}{d^2})^{1/\gamma}\right \rceil $ tests, and an affiliated reconstruction algorithm of computational complexity ${\cal O}\left (\frac{d^2}{\epsilon}\log\left (\frac{n\epsilon}{d^2} \right)\right )$, that ensures a reconstruction error of at most $\epsilon$.
\end{theorem}

\noindent {\bf \underline{Noisy tests:} } 
Finally, we consider the case where there is {\it Bernoulli($\sigma$) noise} in test outcomes -- individual test outcomes are {mis-reported} in an i.i.d. manner, with the probability of misreporting $\sigma \in (0,1/2)$. {In this scenario we show that  when items may be tested at most $\gamma$ times, if $\log(d)/\gamma \in \Omega(1)$, (for instance when $\gamma \in \Theta(1)$, or when $\gamma \in o(\log(n))$ and $d \in \Theta(n^\alpha)$ for some $\alpha$ in $(0,1)$) the probability of error of {\it any} non-adaptive design/reconstruction algorithm is bounded away from zero, {\it regardless} of the number of tests performed.} This is in contrast to results in unconstrained group-testing, wherein reliable reconstruction is still possible, albeit at the cost of a constant (dependent on $\sigma$) factor increase in the number of tests required.

\begin{theorem}[Section~\ref{sec:noise}]
\label{thm:col:noisy}
If non-adaptive test outcomes are corrupted by Bernoulli($\sigma$) noise, $\sigma \in (0,1/2)$, any non-adaptive group-testing algorithm that tests each item at most $\gamma$ times has a probability of error of at least $d\left (\frac{\sigma}{1-\sigma} \right )^\gamma$.\label{thm:column-noise}
\end{theorem}

\subsection{$\rho$-sized tests}
We now discuss our results for the scenario wherein each test may comprise of at most $\rho$ items.

\subsubsection{Parameter regime of interest for $\rho$-sized tests}\label{subsec:row-para}
We begin by again first identifying the parameter regimes of interest for $\rho$. 
Recall that it is well-known in the classical group-testing literature (see for instance~\cite{ChanJSA14,ScarlettC16}) that in unconstrained settings $T >(1-\epsilon)(d\log(n/d))$ tests are necessary to reconstruct all defectives with error probability at most $\epsilon$, hence the same is certainly true in the constrained setting. It is also known (see for instance~\cite{ChanJSA14,ScarlettC16}) that if each test can comprise of $\Theta \left (\frac{n}{d}\right)$ items, then $\Theta(d\log(n))$ tests suffice for group-testing algorithms with vanishing error.
Hence, the parameter regime of primary interest is when $\rho \in o\left (\frac{n}{d}\right)$.
For this $\rho$-sized test model, since our results depend on the ratio between $\log\left (\frac{n}{d}\right)$ and $\log\left (\frac{n}{\rho d}\right)$, it will help to parametrize $d $ as $ \Theta(n^\alpha)$ for some $\alpha \in [0,1)$, and 
$\rho $ as $ \Theta((n/d)^{\beta}) \in \Theta(n^{(1-\alpha)\beta})$ for some $\beta \in [0,1)$. If $d$ or $\rho$ behave like constants, or say logarithmically in $n$, one may then set the corresponding value of $\alpha$ or $\beta$ to equal zero.

To trade off between precision and readability, we specify explicit (upper and lower) bounds in the following theorems, but use instead Bachmann-Landau asymptotic notation to specify parameter regimes and computational complexity.

\subsubsection{Results for $\rho$-sized tests}\label{subsec:row-results}
We now discuss our results for the scenario wherein each test may comprise of at most $\rho$ items. The results here are broadly similar in flavor to those corresponding to the $\gamma$-divisible constrained ones above, with one notable exception being the difference between Theorem~\ref{thm:column-noise} above and Theorem~\ref{thm:row-noise} below.\\

The lower bound we derive in Theorem~\ref{thm:row-lb} for the $\rho$-sized scenario is broadly similar in flavor to the one derived in Theorem~\ref{thm:col-lb}.
\begin{theorem}[Section~\ref{subsec:row-lb}]
\label{thm:row-lb}
For any sufficiently large $n$, sufficiently small $\epsilon > 0$, and $d \in \Theta(n^\alpha)$ for some $\alpha \in [0, 1)$, any non-adaptive group-testing algorithm that includes $\rho \in \Theta((n/d)^{\beta} )$ (for some $\beta \in [0,1)$) items per test and has a probability of error of at most $\epsilon$ requires at least $\left(\frac{1-6\epsilon}{1-\beta}\right )\frac{n}{\rho}$ tests.
\end{theorem}

Analogously to the $\gamma$-divisible group testing schemes in Section~\ref{subsec:col-results}, we now give two different constructions of test matrices (and affiliated computationally tractable reconstruction algorithms) which are row-constrained.

First we give (in Theorem~\ref{thm:row-ub-random}) a construction of a randomized test matrix that requires a number of tests that is larger by a {\it constant} factor of essentially $1/(1-\alpha)$ (neglecting mild dependencies on the probability of reconstruction error) than the lower bound (in Theorem~\ref{thm:row-lb}) to reliably identify $d \in o(n^{\alpha})$ defective items. We note that this additional factor of $1/(1-\alpha)$ is similar to the state of affairs in classical group testing scenario (with no row/column constraints) (for instance see~\cite{ChanJSA14,ScarlettC16}).

For ease of presentation in Theorem~\ref{thm:row-ub-random} below it will be convenient to scale the probability of error $\epsilon$ as $n^{-\zeta}$ for some sufficiently small $\zeta>0$.\footnote{Indeed, it is possible to obtain a similar polynomial decay in the probability of error in each of our achievability algorithms, but to reduce notational clutter we have chosen to specify such scaling for $\epsilon$ only when necessitated by proof details, as in Theorem~\ref{thm:row-ub-random}.}

\begin{theorem}[Section~\ref{sec:row:random}]
\label{thm:row-ub-random}
For any sufficiently large $n$, sufficiently small $\zeta>0$, and $d \in \Theta(n^\alpha)$ for some positive constant $\alpha < 1$, there exists a randomized non-adaptive group-testing design that includes at most $\rho \in \Theta\left (\frac{n}{d} \right )^{\beta}$ (for some positive constant $\beta < 1$) items per test, using at most $\left \lceil \frac{1+\zeta}{(1-\alpha)(1-\beta)} \right \rceil\left \lceil \frac{n}{\rho}\right \rceil$ tests, and an affiliated reconstruction algorithm of computational complexity ${\cal O}\left ( n^2/\rho \right)$, that ensures a reconstruction error of at most $\epsilon = n^{-\zeta}$.
\end{theorem}

In fact, if the scheme in Theorem~\ref{thm:row-ub-random} is used to design a test matrix with $\omega(n/\rho)$ tests instead of $\Theta(n/\rho)$ tests, we then have a {\it universal} design, which works regardless of the value of $d$, as long as $d \in o(n/\rho)$. 
See Remark~\ref{rem:universal} for details.

Our second design results in an explicit construction, analogous to the design in Theorem~\ref{thm:col-ub-explicit}. Even though the number of tests required by the design in Theorem~\ref{thm:row-ub-explicit} is somewhat larger (by a factor of between ${\cal O}(\log(\rho))$ to ${\cal O}(d)$ larger, depending on the regime of $\rho$ and $d$) than the randomized construction, the reconstruction algorithm in Theorem~\ref{thm:row-ub-explicit} may be significantly faster.
To do this, it will help to subdivide $\rho$ into two separate regimes (our design will have different behaviors for these parameter regimes): 
\begin{enumerate}
	\item The first ``test-size constrained'' regime when $\rho$ is less than $\frac{n\epsilon}{d^2}$, and
	\item The second ``block-number constrained'' when $\rho$ is larger than $\frac{n\epsilon}{d^2}$ (but still asymptotically smaller than $\frac{n}{d}$). 
\end{enumerate}

\begin{theorem}[Section~\ref{sec:row:exp}]
\label{thm:row-ub-explicit}
For any sufficiently large $n$, sufficiently small $\epsilon>0$, $d \in \Theta(n^{\alpha})$ for any $\alpha \in [0, 1)$ and $\rho \in {\Theta}(n/d)^{\beta}$ for some $\beta \in [0, 1)$, there exists a deterministic design that ensures a reconstruction error of at most $\epsilon$ such that
\begin{enumerate}
	\item
	When $\rho < \frac{n\epsilon}{d^2}$, the number of tests and computational complexity of reconstruction are both at most $\left \lceil \frac{n}{\rho} \right \rceil \lceil \log(\rho+1)\rceil $.
	\item 
	When $\rho \geq \frac{n\epsilon}{d^2}$, the number of tests and computational complexity of reconstruction are both at most $\left \lceil \frac{d^2}{\epsilon} \right \rceil \lceil \log \left (\frac{n\epsilon}{d^2}+1 \right )\rceil $.
\end{enumerate}
\end{theorem}

\noindent {\bf \underline{Noisy tests:} }
Finally, we consider the impact of noise in test outcomes when the size of each test is constrained to at most $\rho$. 
The situation here is more akin to the situation in unconstrained noisy non-adaptive group-testing (for instance see~\cite{atia2012boolean,scarlett2017limits,ChanJSA14}) wherein one can ensure reliable reconstruction even from noisy test outcomes at the cost of at most a ``small'' multiplicative factor in the number of test outcomes -- a constant factor (dependent on $\sigma$ but independent of $d$ and $n$) in unconstrained group-testing. 
In the setting we now consider a factor that is at most logarithmic in $n$ will suffice to ensure reliable recovery (i.e., $\Theta(\frac{n\log(n)}{\rho})$ tests suffice in this setting), and indeed, as discussed in Remark~\ref{rem:nolog}, it is conceivable that even the need for this logarithmic factor can be obviated by somewhat more sophisticated arguments than those we consider in this work. 
\begin{theorem}[Section~\ref{sec:noise}]
\label{thm:row-noise}
For any sufficiently large $n$, sufficiently small $\zeta>0$, and $d \in \Theta(n^\alpha)$ for some positive constant $\alpha < 1$, there exist a randomized non-adaptive group-testing design that includes at most $\rho \in \Theta\left (\frac{n}{d} \right )^{\beta}$ (for some positive constant $\beta < 1$) items per test, using at most $\left \lceil \frac{1+\zeta}{(1-\alpha)(1-\beta)} \right \rceil\left \lceil \frac{n}{\rho}\right \rceil \left \lceil \frac{(1+\zeta)\ln(n)}{(1/2-\sigma)^2}\right \rceil$ tests, and an affiliated reconstruction algorithm of computational complexity ${\cal O}\left ( \frac{n^2\log(n)}{\rho} \right)$, that ensures a reconstruction error of at most $\epsilon = 2n^{-\zeta}$.
\end{theorem}

All results stated in this section are summarized in Table~\ref{table:results}

\begin{table}[H]
\centering
\resizebox{0.8\columnwidth}{!}
{
\input{table}
}
\caption{A summary of non-adaptive group testing results.}
\label{table:results}
\end{table}

%% file: table.tex
\begin{tabular}{| c | c | c | c | c |}\hline

\multicolumn{2}{|c|}{\multirow{2}{*}{Model}}&
\multirow{2}{*}{Regime}&
\multirow{2}{*}{{\small Tests}}&
\multirow{2}{*}{{\small Computational Complexity}} \\
\multicolumn{2}{|c|}{} &&&\\\hline\hline

\parbox[t]{2mm}{\multirow{18}{*}{\rotatebox[origin=c]{90}{{\small General}}}}& 
\parbox[t]{2mm}{\multirow{6}{*}{\rotatebox[origin=c]{90}{{\small Randomized}}}}&
\multirow{2}{*}{$\gamma \in \Theta(\log(n))$ }& 
& 
\multirow{6}{*}{}\\ 
&&&$T>(1-\epsilon)d\log(n/d)$~ \cite{ChanJSA14}&\\

& 
& 
\multirow{2}{*}{$\rho\in \Theta(n/d)$}& 
$T < (1+\epsilon)d\log(n/d)$ for $d \in o(n^{1/3})$ ~\cite{ScarlettC16}& 
${\cal O}(n^{d})$ \\ 
&&&$T < {\cal O}((1+\epsilon)d\log(n/d))$~\cite{johnson2018performance}&${\cal O}(dn\log(n))$\\

&&\multirow{2}{*}{$d=o(n^{\alpha}), \alpha<1$}&$T <  {\cal O}( \log(d)(1+\epsilon)d\log(n/d))$~\cite{lee2016saffron}&${\cal O}(d\log(d)\log(n)$\\
&&&&\\\cline{2-5}

&  
\parbox[t]{2mm}{\multirow{6}{*}{\rotatebox[origin=c]{90}{{\small Explicit}}}}& 
\multirow{6}{*}{Same as above}& 
\multirow{3}{*}{Same as randomized}& 
\multirow{6}{*}{}\\ 
&&&&\\
&&&&\\

& 
& 
& 
\multirow{3}{*}{$T < O\left(d \frac{\log n}{\log d} \log(\frac{n}{\epsilon})\right)$~\cite{Mazumdar15}} & ${\cal O}\left( dn \log n \right)$ 
\\ 

&&&&\\
&&&&\\\cline{2-5}

& 
\parbox[t]{2mm}{\multirow{6}{*}{\rotatebox[origin=c]{90}{{\small Noisy}}}}& 
\multirow{6}{*}{Same as above}& 
\multirow{6}{*}{$T < {\cal O}\left( d \log(\frac nd)\right) $ ~~~\cite{scarlett2018near}}& 
\multirow{6}{*}{${\cal O}\left( d^2 \log^2(\frac nd)\right)$}\\ 
&&&&\\
&&&&\\

& 
& 
& 
& 
\\

&&&&\\
&&&&\\\cline{2-5}
\hline\hline

\parbox[t]{2mm}{\multirow{18}{*}{\rotatebox[origin=c]{90}{{\small $\gamma$-divisible items}}}}& 
\parbox[t]{2mm}{\multirow{6}{*}{\rotatebox[origin=c]{90}{{\small Randomized}}}}&
\multirow{3}{*}{$d = o(n^{\alpha})$, $\alpha < 1$}& 
\multirow{3}{*}{$T > \gamma d\left(\frac{n}{d}\right)^{\frac{1-6\epsilon}{\gamma}}$~~~[Thm~\ref{thm:col-lb}]}& 
\multirow{6}{*}{${\cal O}\left ( \gamma d\left(\frac{n}{\epsilon}\right)^{1+1/\gamma}\right )$}\\ 
&&&&\\
&&&&\\

& 
& 
\multirow{3}{*}{$\gamma=o(\log n)$}& 
\multirow{3}{*}{$T < \left\lceil e\gamma d\left(\frac{n}{\epsilon}\right)^{\frac{1}{\gamma}}\right\rceil$~~~[Thm~\ref{thm:col-ub-random}]}& 
\\ 

&&&&\\
&&&&\\\cline{2-5}

&  
\parbox[t]{2mm}{\multirow{6}{*}{\rotatebox[origin=c]{90}{{\small Explicit}}}}& 
\multirow{3}{*}{$d = o(n^{\alpha})$, $\alpha < \frac12$}& 
\multirow{3}{*}{Same as Randomized}& 
\multirow{6}{*}{${\cal O}\left (\frac{d^2}{\epsilon}\log\left (\frac{n\epsilon}{d^2} \right)\right )$}\\ 
&&&&\\
&&&&\\

& 
& 
\multirow{3}{*}{$\gamma=o(\log n)$}& 
\multirow{3}{*}{$T < \lceil\frac{d^2\gamma}{\epsilon}\rceil\left \lceil (\frac{n\epsilon}{d^2})^{1/\gamma}\right \rceil $~~~[Thm~\ref{thm:col-ub-explicit}]}& 
\\ 

&&&&\\
&&&&\\\cline{2-5}

& 
\parbox[t]{2mm}{\multirow{6}{*}{\rotatebox[origin=c]{90}{{\small Noisy}}}}& 
\multirow{3}{*}{Noise: Bernoulli($\sigma$), $\sigma \in (0,1/2)$ }& 
\multirow{6}{*}{Impossible~~~[Thm~\ref{thm:col:noisy}]}& 
\multirow{6}{*}{}\\ 
&&&&\\
&&&&\\

& 
& 
\multirow{3}{*}{$\log(d)/\gamma \in \Omega(1)$}& 
& 
\\ 

&&&&\\
&&&&\\\cline{2-5}
\hline\hline

\parbox[t]{2mm}{\multirow{18}{*}{\rotatebox[origin=c]{90}{{\small $\rho$-sized tests}}}}& 
\parbox[t]{2mm}{\multirow{6}{*}{\rotatebox[origin=c]{90}{{\small Randomized}}}}&
\multirow{2}{*}{$d \in \Theta(n^\alpha)$, $\alpha<1$}& 
\multirow{3}{*}{$T > \left(\frac{1-6\epsilon}{1-\beta}\right )\frac{n}{\rho}$~~~[Thm~\ref{thm:row-lb}]}& 
\multirow{6}{*}{${\cal O}\left(\frac{n}{\rho}\log\left(\frac{n}{\epsilon}\right)\right)$}\\ 
&&&&\\

& 
& 
\multirow{2}{*}{$\rho \in \Theta((n/d)^{\beta} )$, $\beta < 1$}& 
\multirow{3}{*}{$T < \left \lceil \frac{1+\zeta}{(1-\alpha)(1-\beta)} \right \rceil\left \lceil \frac{n}{\rho}\right \rceil$~~~[Thm~\ref{thm:row-ub-random}]}& 
\\ 

&&&&\\
&&\multirow{2}{*}{$\epsilon = n^{-\zeta}$, $\zeta>0$}&&\\
&&&&\\\cline{2-5}

&  
\parbox[t]{2mm}{\multirow{6}{*}{\rotatebox[origin=c]{90}{{\small Explicit}}}}& 
\multirow{3}{*}{$d \in \Theta(n^{\alpha})$, $\alpha<1$}& 
\multirow{3}{*}{Same as randomized}& 
\multirow{6}{*}{${\cal O}(T)$}\\ 
&&&&\\
&&&&\\

& 
& 
\multirow{3}{*}{$\rho \in {\Theta}(n/d)^{\beta}$, $\beta <1$}& 
\multirow{3}{*}{$T <
     \begin{cases}
   	\left \lceil \frac{n}{\rho} \right \rceil \lceil \log(\rho+1)\rceil  &  \mbox{if } \rho > \frac{n\epsilon}{d^2}\\
	\left \lceil \frac{d^2}{\epsilon} \right \rceil \lceil \log \left (\frac{n\epsilon}{d^2}+1 \right )\rceil  &  \mbox{ if } \rho < \frac{n\epsilon}{d^2}
     \end{cases}
$ ~~~[Thm~\ref{thm:row-ub-explicit}]}& 
\\ 

&&&&\\
&&&&\\\cline{2-5}

& 
\parbox[t]{2mm}{\multirow{6}{*}{\rotatebox[origin=c]{90}{{\small Noisy}}}}& 
\multirow{2}{*}{$d \in \Theta(n^\alpha)$, $\alpha<1$}& 
\multirow{6}{*}{$T < \left \lceil \frac{1+\zeta}{(1-\alpha)(1-\beta)} \right \rceil\left \lceil \frac{n}{\rho}\right \rceil \left \lceil \frac{(1+\zeta)\ln(n)}{(1/2-\sigma)^2}\right \rceil$~~~[Thm~\ref{thm:row-noise}]}& 
\multirow{6}{*}{${\cal O}\left ( \frac{n^2\log(n)}{\rho} \right)$}\\ 
&&&&\\

& 
& 
\multirow{2}{*}{$\rho \in \Theta\left (\frac{n}{d} \right )^{\beta}$, $\beta<1$}& 
& 
\\ 

&&&&\\
&&\multirow{2}{*}{Error: $\epsilon = 2n^{-\zeta}$}&&\\
&&&&\\\cline{2-5}
\hline

\end{tabular}

%% file: column-information.tex
\section{$\gamma$-Divisible Items}\label{sec:gamma-div}
In Section~\ref{sec:col:lb} we present the proof of an information-theoretic lower bound on the number of tests required by any non-adaptive group-testing scheme that is allowed to test each item no more than $\gamma$ times and has probability of error no more than $\epsilon$. 
In Section~\ref{sec:col:random} we provide a randomized construction of a corresponding group-testing algorithm, and in Section~\ref{sec:col:exp} we provide an alternative explicit construction (that requires more tests than the randomized construction, but on the other hand has significantly smaller computational complexity of  decoding).

\subsection{Proof of Theorem~\ref{thm:col-lb}: Information-Theoretic Lower Bounds}
\label{sec:col:lb}
We begin by partitioning the tests $T$ into sets $S_l$ and $S_h$, where $i\in S_l$ if test $i$ includes less than $\frac{n}{\epsilon d\log\left(\frac{T}{\gamma d}\right)}$ items, and $i\in S_h$ otherwise (that is, test $i$ includes at least $\frac{n}{\epsilon d\log\left(\frac{T}{\gamma d}\right)}$ items).
Roughly speaking, tests in set $S_l$ are ``light'' (test ``few'' items per test) and hence have a ``high'' probability of being negative, and thus ``low'' entropy (significantly less than $1$ bit per test). 
Conversely, tests in set $S_h$ are ``heavy'' (test ``many items per test) and may potentially have ``high'' entropy  (as much as $1$ bit per test) -- however, there cannot be too many heavy tests, due to the constraint that each item is tested at most $\gamma$ times.

We first bound the entropy of the heavy tests. 
Since there are at most a total of $\gamma n$ $1$'s in the test matrix and the entropy of each test outcome binary variable $Y_i$ is at most $1$, then
\begin{equation}
\sum_{i\in S_h}H(Y_i)\le|S_h|\le\frac{\gamma n}{\left (\frac{n}{\epsilon d\log\left(\frac{T}{\gamma d}\right)}\right)}=\epsilon\gamma d\log\left(\frac{T}{\gamma d}\right).\label{eqn:s2}	
\end{equation}

Next, we bound from above the entropy of the light tests. For any $i \in T$, let $g_i$ denote the number of items in test $i$. We first note that since for light tests, $g_i \leq \frac{n}{\epsilon d\log\left(\frac{T}{\gamma d}\right)}$, hence $\frac{dg_i}{n} \leq \frac{1}{\epsilon \log\left(\frac{T}{\gamma d}\right)}$, and hence by 
Equation~\eqref{eq:Tgammad},
\begin{equation}
\frac{dg_i}{n} \in o(1)\label{eq:dgin}
\end{equation}

We next bound from below, for any fixed test $i$ with a fixed set of $g_i$ items being tested, the probability (with randomness uniformly distributed over all possible ${\dbinom{n}{d}}$ sets of defectives) that test outcome $Y_i$ is negative, and denote this probability $p^-_i$. While the precise value of $p^-_i$ equals $\frac{\dbinom{n-g_i}{d}}{\dbinom{n}{d}}$, via some direct calculations this equals 
$\frac{(n-g_i)!(n-d)!}{n!(n-d-g_i)!} = \frac{\dbinom{n-d}{g_i}}{\dbinom{n}{g_i}}$, which may be interpreted as the probability that for a {\it fixed set of $d$ items} being defective, the probability (with randomness uniformly distributed over all possible ${\dbinom{n}{g_i}}$ sets of {\it tests of size $g_i$}) that test outcome $Y_i$ is negative, i.e., the randomness is shifted from sampling the defectives without replacement $d$ times, to sampling the test items without replacement $g_i$ times. But for a fixed set of $d$ defectives the probability of a negative test outcome  by sampling $g_i$ items {\it without} replacement is strictly greater than that of sampling $g_i$ items {\it with} replacement, which in turn equals $\left (1-\frac{d}{n}\right )^{g_i}$.
Hence
\begin{equation}
p^-_i > \left (1-\frac{d}{n}\right )^{g_i} \geq 1-\frac{dg_i}{n} \label{eq:berns},
\end{equation}
where the latter inequality inequality follows from Bernoulli's identity~\cite{samuel1997advanced}.

Hence, for light tests, the probability of negative test outcomes is very close to $1$. This allows us to bound the entropy of any individual light test from above as follows:
\begin{align}
H(Y_i) &= H(p_i^-) = H(1-p^-_i) < H \left (\frac{dg_i}{n} \right )\nonumber\\
&= \frac{dg_i}{n}\log \left (\frac{n}{dg_i} \right ) - \left(1-\frac{dg_i}{n} \right )\log \left (1 - \frac{dg_i}{n} \right )\nonumber\\
& \leq  \frac{dg_i}{n}\log \left (\frac{n}{dg_i} \right ) + \frac{dg_i}{n} \label{eq:ent_tay} \\
&\leq \frac{dg_i}{n}\log \left (\frac{n}{dg_i} \right )(1+\epsilon).\nonumber
\end{align}
for any $\epsilon >0$ and sufficiently large $n$. 
Here Equation~\eqref{eq:ent_tay} follows using the fact that $-\log(1-x)\leq \frac{x}{1-x}$, for $x<1$, $x\ne 0$.

Note that since the total number of $1$'s in the test matrix is at most $\gamma n$, the entropy over all light tests ($\sum_{i \in S_l} H(Y_i)$) is bounded from above by the following constrained optimization problem, where the optimization variables are the set of $g_i$ for light tests: 
\begin{eqnarray}
(1+\epsilon)\max_{\{g_i:i\in S_l\}}\left (\sum_{i \in S_l} \frac{dg_i}{n}\log \left (\frac{n}{dg_i} \right ) \right ) \mbox{ subject to }  \sum_{i \in S_l} g_i \leq \gamma n.
\label{eq:optim}
\end{eqnarray}

It can be readily verified (via, for instance the method of Lagrange multipliers) that the maximum of~\eqref{eq:optim} occurs when each of the $g_i$'s are equal, and hence each is at most $\frac{\gamma n}{|S_l|}$. 
Thus, since $|S_l| \leq T$, the total entropy over all the light tests is at most
\begin{equation}
	|S_l|(1+\epsilon)\left ( \frac{d\gamma}{|S_l|}\log \left (\frac{|S_l|}{d\gamma} \right ) \right ) \leq (1+\epsilon){d\gamma}\log \left (\frac{T}{d\gamma} \right ). \label{ent_heavy}
\end{equation}

Hence, adding Equation~\eqref{eqn:s2} to \eqref{ent_heavy}, the overall entropy $H(Y)$ of all test outcomes, is bounded from above by
\begin{equation}
H(Y) \leq \sum_{i=1}^T H(Y_i)\le (1+2\epsilon){d\gamma}\log \left (\frac{T}{d\gamma} \right ).\label{eqn:tot_ent}	
\end{equation}
The remainder of this proof follows relatively standard lines (see for instance~\cite{atia2012boolean,ChanJSA14}) in the literature on information-theoretic converses for (classical) group-testing. 
Specifically, we begin by noting that $X\leftrightarrow Y\leftrightarrow\widehat{X}$ form a Markov chain.\footnote{Note that this implicitly assumes that the specific test-matrix $M$ is deterministically fixed in advance. 
As is standard in information-theoretic this is without loss of generality in the average probability of error setting.}
From standard information-theoretic definitions, we have 
\begin{equation}\label{eqn:e1}
H(X)=H(X|\widehat{X})+I(X;\widehat{X}),
\end{equation}
where $H(X)$ is the binary entropy of the length-$n$ binary vector $X$, and $I(X;\widehat{X})$ is the mutual information between $X$ and $\widehat{X}$. 
Since $X$ is uniformly distributed over $\mathcal{X}$, the set of all length-$n$, $d$-sparse binary vectors, we have
\begin{equation}\label{eqn:main}
H(X)=\log|\mathcal{X}|=\log\binom{n}{d}.
\end{equation}

\noindent
We now upper bound each of the terms in RHS of Equation~\ref{eqn:e1} separately. 
By Fano's Inequality, 
\begin{equation}H(X|\widehat{X})\le H(\epsilon)+\epsilon\log(|\mathcal{X}|-1).
\label{eq:fano}	
\end{equation}

\noindent
Note that for $\epsilon<\frac{1}{2}$, 
\begin{equation}\label{eqn:fanos}
H(\epsilon)<-2\epsilon\log\epsilon.
\end{equation}
Also, by the data processing inequality and standard information theoretic inequalities, 
\[I(X;\widehat{X})\le I(X;Y)=H(Y)-H(Y|X)\le H(Y).\] 
Combining Equations~\eqref{eqn:main}, \eqref{eqn:fanos}, \eqref{eq:fano}, \eqref{eqn:tot_ent} and~\eqref{eqn:e1} we have
\begingroup
\allowdisplaybreaks
\begin{align*}
&&H(X) &=H(X|\widehat{X})+I(X;\widehat{X})\le H(\epsilon)+\epsilon\log(|\mathcal{X}|-1) + H(Y) \\
&\Rightarrow &\log\binom{n}{d} & \le -2\epsilon\log\epsilon + \epsilon\log\binom{n}{d} + (1+2\epsilon)\gamma d\log\left(\frac{T}{\gamma d}\right).
\end{align*}
\endgroup
By reordering the terms we get a lower bound on the number of tests as
\begingroup
\allowdisplaybreaks
\begin{align*}
T &\ge \gamma d \exp\left({\frac{(1-\epsilon)\log\binom{n}{d}+2\epsilon\log\epsilon}{(1+2\epsilon)\gamma d}}\right)\\
& \ge \gamma d \exp\left({\frac{(1-5\epsilon)\log\binom{n}{d}}{\gamma d}}\right) &\text{(for sufficiently large $n$)}\\
&=\gamma d\binom{n}{d}^{\frac{1-5\epsilon}{\gamma d}} \ge \gamma d\left(\frac{n}{d}\right)^{\frac{1-5\epsilon}{\gamma d}}&\text{(by Fact~\ref{fact:stirling})}.
\end{align*}
\endgroup
Hence, $T \geq \gamma d\left(\frac{n}{d}\right)^{\frac{1-5\epsilon}{\gamma d}}$ tests are needed by any non-adaptive group-testing procedure that has probability of error at most $\epsilon$. 

\begin{remark}
As pointed out by an anonymous reviewer, the approach followed in the above proof does not carry through for adaptive group testing, since our proof bounding the entropy $H(Y_i)$ of individual testing outcomes critically relies on the test matrix $M$ being independent of $X$. 
Since the focus of this work is on the nonadaptive setting, we leave open the question of deriving lower bounds for the adaptive setting as an interesting open question.
\end{remark}

%% file: column-construction.tex
\subsection{Proof of Theorem~\ref{thm:col-ub-random}: Randomized Construction of Test Matrices}
\label{sec:col:random}
We now describe a randomized construction of a $T\times n$ test matrix $M$, where $T=\left \lceil e\gamma d\left(\frac{n}{\epsilon}\right)^{1/\gamma} \right \rceil $ tests suffice to guarantee a probability of reconstruction error of at most $\epsilon$ via a reconstruction algorithm of computational complexity ${\cal O}\left (\gamma d\left(\frac{n}{\epsilon}\right)^{1+1/\gamma}\right )$. 
The test matrix is obtained by picking each column of $M$ uniformly at random from the set $\{0,1\}^T$ of length $T$ binary vectors of Hamming weight $\gamma$. We now describe how to reconstruct the estimate vector $\widehat{X}$ from the test results. 
This reconstruction algorithm is essentially the same as one outlined for classical group-testing in~\cite{ChanJSA14}, adapted to the parameter settings in this paper.\\

\subsubsection{The Column Matching Algorithm (CoMa)}\label{subsubsec:coma}
To obtain the estimate vector $\widehat{X}$ from result vector $Y$, the Column Matching algorithm (CoMa) from \cite{ChanJSA14} uses the tests which have positive outcomes to identify all defective items, while declaring all other items to be non-defective. 
Namely, the algorithm marks item $i$ defective if every test in which $i$ is included is positive. 
Note that CoMa cannot incorrectly mark defective items as non-defective. 
CoMa can only incorrectly designate a non-defective item as defective if the item is not tested, or is only tested in positive tests (\emph{i.e.,} every test it occurs in has at least one defective item). 
If $M$ is chosen to have enough rows, then we show that with significant probability, each non-defective item appears in at least one negative test, and hence will be appropriately marked non-defective. 
Note that the computational complexity of this reconstruction algorithm is therefore $Tn = {\cal O} \left (\gamma d\left(\frac{n}{\epsilon}\right)^{1+1/\gamma} \right ).$

\subsubsection{Analysis}
Since each of the $d$ defective items can be tested at most $\gamma$ times, the maximum number of tests which are positive is at most $d\gamma$.
Now, an item will be marked by CoMa as defective if all the tests which pick this particular item are positive.
Therefore for a fixed non-defective item $i$, the probability that it is incorrectly marked defective is the probability that $i$ is always tested with one of the $d$ defective items which happens with probability at most $\binom{d\gamma}{\gamma}/\binom{T}{\gamma}$. 
Taking a union bound over the $(n-d)$ non-defective items, we require $(n-d)\binom{d\gamma}{\gamma}/\binom{T}{\gamma}\leq \epsilon$. 
From Fact~\ref{fact:stirling}, we know that this condition is satisfied if $\binom{d\gamma}{\gamma}<(ed)^\gamma$ and $\binom{T}{\gamma}>\left(\frac{T}{\gamma}\right)^\gamma$ then this certainly occurs if $(ed)^\gamma(n-d)\leq \epsilon\left(\frac{T}{\gamma}\right)^\gamma$ (since $\binom{d\gamma}{\gamma}<(ed)^\gamma$ and $\binom{T}{\gamma}>\left(\frac{T}{\gamma}\right)^\gamma$). 
Thus, we see that choosing $T$ as $ \left \lceil e\gamma d\left(\frac{n}{\epsilon}\right)^{1/\gamma}\right \rceil$ suffices to ensure correct recovery of the set of defective items with a probability of error of at most $\epsilon$.

\begin{remark}\label{rem:col:ratio}
Note the ratio between the number of tests required via the algorithm considered here and the lower bound in Section~\ref{sec:col:lb} scales essentially as $\Theta((d/\epsilon)^{1/\gamma})$ (neglecting lower-order dependencies on the probability of error $\epsilon$), which for large values of $d$ (say $d$ scaling as $n^{\alpha}$ for some $\alpha \in (0,1)$) may be significant. 
It is conceivable that there is room to improve on our upper bound. 
Specifically, if one were to consider the Definitely Defective decoder (see, for instance,~\cite{AldridgeBJ}) or Maximum Likelihood decoding (for instance, via the approach followed in~\cite{ScarlettC16}) instead of the greedy CoMa decoding considered in this section, it is possible that one may be able to significantly improve the gap between the upper and lower bounds in this work. 
Alternatively, it might also be possible to improve our lower bound in this model, by using the approach in~\cite{agarwal2018novel} (attempting to quantify the correlation between tests, and thereby obtaining a tighter bound on $H(Y)$ than $\sum_{i=1}^T H(Y_i)$), or the approach in~\cite{aldridge2018individual} (which bounds from below the probability that at least one item is completely masked by other items, and hence leads to error). 
We leave these directions open as interesting questions to be explored in future work.
\end{remark}

%% file: column-explicit.tex
\subsection{Proof of Theorem~\ref{thm:col-ub-explicit}: Explicit Construction of Test Matrices}
\label{sec:col:exp}
In this section we focus on explicit constructions of non-adaptive test matrices. 
Even though the explicit construction requires more tests than the randomized construction (shown in Section~\ref{sec:col:random}), its decoding complexity far better. 
We first attempt to generalize the grid construction for $\gamma=2$ in Section~\ref{sec:proof-outline}, and point out a shortcoming in a na\"ive implementation.

\subsubsection{First Tool: $\gamma$-Dimensional Hypergrid}\label{subsubsec:hypergrid}
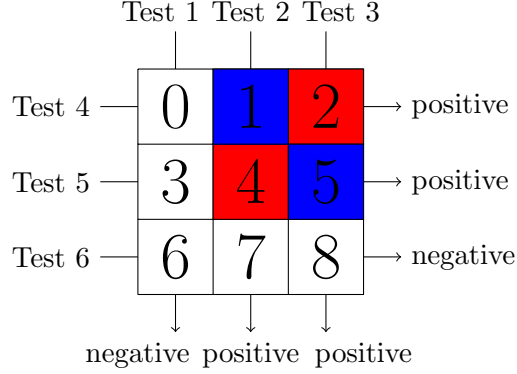
\begin{figure}[!ht]
\centering
\begin{tikzpicture}[scale=1]
\draw[step=1cm] (0,0) grid (3,3);
\draw [fill=red] (1cm,1cm) rectangle (2cm,2cm);
\draw [fill=red] (2cm,2cm) rectangle (3cm,3cm);
\draw [fill=blue] (2cm,1cm) rectangle (3cm,2cm);
\draw [fill=blue] (1cm,2cm) rectangle (2cm,3cm);
\node at (0.5cm,0.5cm){\Huge 6};
\node at (1.5cm,0.5cm){\Huge 7};
\node at (2.5cm,0.5cm){\Huge 8};
\node at (0.5cm,1.5cm){\Huge 3};
\node at (1.5cm,1.5cm){\Huge 4};
\node at (2.5cm,1.5cm){\Huge 5};
\node at (0.5cm,2.5cm){\Huge 0};
\node at (1.5cm,2.5cm){\Huge 1};
\node at (2.5cm,2.5cm){\Huge 2};
\draw (0.5cm,3.5cm) -- (0.5cm,3cm);
\draw (1.5cm,3.5cm) -- (1.5cm,3cm);
\draw (2.5cm,3.5cm) -- (2.5cm,3cm);
\draw [->](0.5cm,0cm) -- (0.5cm,-0.5cm);
\draw [->](1.5cm,0cm) -- (1.5cm,-0.5cm);
\draw [->](2.5cm,0cm) -- (2.5cm,-0.5cm);
\node at (0.3cm,3.5cm)[above]{Test 1};
\node at (1.5cm,3.5cm)[above]{Test 2};
\node at (2.7cm,3.5cm)[above]{Test 3};
\draw (-0.5cm,0.5cm) -- (0,0.5cm);
\draw (-0.5cm,1.5cm) -- (0,1.5cm);
\draw (-0.5cm,2.5cm) -- (0,2.5cm);
\draw [->](3cm,0.5cm) -- (3.5cm,0.5cm);
\draw [->](3cm,1.5cm) -- (3.5cm,1.5cm);
\draw [->](3cm,2.5cm) -- (3.5cm,2.5cm);
\node at (-0.5cm,0.5cm)[left]{Test 6};
\node at (-0.5cm,1.5cm)[left]{Test 5};
\node at (-0.5cm,2.5cm)[left]{Test 4};
\node at (0cm,-0.5cm)[below]{negative};
\node at (1.5cm,-0.5cm)[below]{positive};
\node at (3cm,-0.5cm)[below]{positive};
\node at (3.5cm,0.5cm)[right]{negative};
\node at (3.5cm,1.5cm)[right]{positive};
\node at (3.5cm,2.5cm)[right]{positive};
\end{tikzpicture}
{\caption{If $n=9,\gamma=2,d=2$, the above test cannot distinguish whether the red items or the blue items are defective. However, if there were only one defective item, the series of tests would uniquely identify the defective item.}\label{fig:grid}}
\end{figure}

For ease of presentation, define $b=\lceil n^{1/\gamma}\rceil $. 
We represent each item $i\in\{0,\ldots,n-1\}$ by its base-$b$ representation $(x_\gamma\ldots x_2x_1)_b$, so that each $x_j\in\{0,1,\ldots,b-1\}$ and 
\[i=\sum_{j=1}^\gamma x_jb^{j-1}.\]
For test $t$, where $t=\alpha b+k$, for $\alpha\in\{0,1,\ldots,\gamma-1\}$ and $k\in\{0,1,\ldots,b-1\}$, we include exactly the items whose $(\alpha+1)$th coordinate is $k$, \emph{i.e.}, $x_{\alpha+1}=k$.
Hence, there are $\gamma b=\gamma \left \lceil n^{1/\gamma}\right \rceil$ tests in total. 
See Figure \ref{fig:grid} for an example.

Intuitively, test $t=\alpha b+k$ returns whether or not there exists a defective item $i$ whose base-$b$ representation has $x_{\alpha+1}=k$. 
Note that a defective item $i\in[n]$ will cause exactly $\gamma$ tests to be positive, corresponding to when each of its coordinates is tested. 
Thus, if there exists a unique defective item, it can be successfully recovered from its unique base-$b$ representation.

\begin{remark}\label{rem:compcomhypergrid}
Indeed, note that if the vector $Y$ of test outcomes is provided as a list of indices corresponding to positive test outcomes (rather than the full length-$T$ vector $Y$), then the computational complexity of the reconstruction algorithm is just ${\cal O}(\gamma \log(n^{1/\gamma})) = {\cal O}(\log(n))$.
\end{remark}

However, with multiple defective items, we may not be able to uniquely determine each item. For example, for $n=9$, $d=2$ and $\gamma=2$, if items $2$ and $4$ are defective, then positive tests will tell us that there exist defective items with $x_1=1$ (corresponding to item $4$), $x_1=2$ (corresponding to item $2$), $x_2=0$ (corresponding to item $2$) and $x_2=1$ (corresponding to item $4$). 
However, another pair of defective items which return the same positive test results are items $1$ and $5$. 
Thus, we cannot uniquely recover all defective items, unless there is only one defective item. 
See Figure \ref{fig:grid} for more details.

\subsubsection{Block Algorithm: Divide and Conquer}
\begin{figure}[!ht]
\centering
\begin{tikzpicture}[scale=0.75]
\draw[step=1cm] (0,0) grid (5cm,5cm);
\filldraw[shading=radial,inner color=white, outer color=gray!75, opacity=0.2](0,5cm) rectangle (1cm,4cm);
\filldraw[shading=radial,inner color=white, outer color=gray!75, opacity=0.2](1cm,4cm) rectangle (2cm,3cm);
\filldraw[shading=radial,inner color=white, outer color=gray!75, opacity=0.2](2cm,3cm) rectangle (3cm,2cm);
\filldraw[shading=radial,inner color=white, outer color=gray!75, opacity=0.2](3cm,2cm) rectangle (4cm,1cm);
\filldraw[shading=radial,inner color=white, outer color=gray!75, opacity=0.2](4cm,1cm) rectangle (5cm,0cm);
\end{tikzpicture}
{\caption{The test matrix for the block algorithm, where each gray block represents the test matrix for the first part.}\label{fig:block}}

\end{figure}
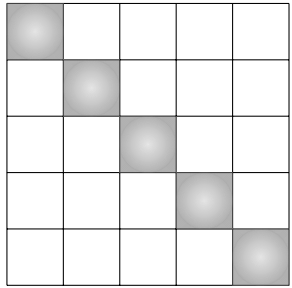

When $d = o(\sqrt{n})$ we now provide an explicit construction of a $T\times n$ test matrix $M$, where $T=\lceil\frac{d^2\gamma}{\epsilon}\rceil\left \lceil (\frac{n\epsilon}{d^2})^{1/\gamma}\right \rceil $, using the previous ideas. 
The key observation is that the algorithm from Section~\ref{subsubsec:hypergrid} succeeds if there is a unique defective item. 
Thus, we split $[n]$ into $\lceil cd^2\rceil$ blocks, where $c=\frac{1}{\epsilon}$, and run the algorithm in Section~\ref{subsubsec:hypergrid} separately on each block of size $n'=n/\lceil cd^2\rceil$. 
(See Figure \ref{fig:block} for an example.) 
Then the probability that no two defective items fall into the same block is at most
\begingroup
\allowdisplaybreaks
\begin{align}
1\left(1-\frac{1}{cd^2}\right)\left(1-\frac{2}{cd^2}\right)\cdots\left(1-\frac{d-1}{cd^2}\right)&\ge\left(1-\frac{d}{cd^2}\right)^d=\left(1-\frac{1}{cd}\right)^d&\label{eq:anti-bday}\\
&\ge1-\frac{1}{c}=1-\epsilon&\text{(by Bernoulli's Inequality)}.\nonumber
\end{align}
\endgroup
Thus with probability at least $1-\epsilon$ no block contains more than one defective item, so we can also successfully identify the $d$ defective items with  probability at least $1-\epsilon$ using the algorithm in Section~\ref{subsubsec:hypergrid} for $n'$ items. 
Since there are at most $\lceil\frac{d^2}{\epsilon}\rceil$ blocks, each requiring at most $\gamma \left \lceil\left(\frac{n\epsilon}{d^2}\right)^{1/\gamma}\right \rceil$ tests, this leads to a total of $T=\lceil\frac{d^2\gamma}{\epsilon}\rceil\left \lceil (\frac{n\epsilon}{d^2})^{1/\gamma}\right \rceil $ tests.

\begin{remark}
\label{remark:cc:explicit}
Building on Remark~\ref{rem:compcomhypergrid}, we note that the computational complexity of the decoding algorithm for this explicit design is ${\cal O}\left (\frac{d^2}{\epsilon}\log\left (\frac{n\epsilon}{d^2} \right)\right )$. 
Hence, while the number of tests required by this explicit design is larger (by a factor of ${\Theta}\left (\frac{d}{\epsilon} \right )^{1-2/\gamma}$) than the randomized design discussed in Section~\ref{subsubsec:coma}, for small values of $d$ (for instance if $d$ is constant or logarithmic in $n$) the computational complexity of the reconstruction algorithm may be {\it exponentially} less than that of the CoMa algorithm in Section~\ref{subsubsec:coma}. 
Explicit constructions (with affiliated fast reconstruction algorithms), perhaps using the coding-theoretic ideas in~\cite{barg2017group,lee2016saffron,cai2017efficient,indyk2010efficiently}, perhaps also obviating the need to restrict $d \in o(\sqrt{n})$ rather than $d \in o(n)$, is an interesting direction for future research.
\end{remark}

%% file: row-information.tex
\section{$\rho$-Sized Tests}\label{sec:row:lb}
\label{sec:gamma-div}
This section parallels the results in Section~\ref{sec:gamma-div}, but with constraints on the size of each test rather than constraints on the number of times items can be divided. 
Specifically, in Section~\ref{subsec:row-lb} we present the proof of an information-theoretic lower bound on the number of tests required, in Section~\ref{sec:row:random} we provide a randomized construction of a corresponding group-testing algorithm, and in Section~\ref{sec:row:exp} we provide an alternative explicit construction.

\subsection{Proof of Theorem~\ref{thm:row-lb}: Information-Theoretic Lower Bounds}\label{subsec:row-lb}
This proof broadly parallels the one in Section~\ref{sec:col:lb} with some simplifications (in particular, one does not have to divide tests into light and heavy tests -- all tests may be treated as light tests, with at most $\rho$ items).

Via the same approach as in Section~\ref{sec:col:lb} (see the argument leading up to Equation~\eqref{eq:berns}), noting that each test may have at most $\rho$ items, the probability of a negative test outcome is bounded from below by $\left(1-\frac{d}{n}\right)^{\rho}\ge 1-\frac{\rho d}{n}$.

Hence, for $d \in o(n^\alpha)$ for some constant $\alpha \in [0,1)$, $\rho \in \Theta(n/d)^{\beta}$ for some constant $\beta  \in [0,1)$, and any sufficiently small positive $\epsilon$, we have that $H(Y_i)\le(1+\epsilon)\left(\frac{\rho d}{n}\log\left(\frac{n}{\rho d}\right)\right)$. 
Therefore,
\begin{align*}
& H(X)=H(X|\widehat{X})+I(X;\widehat{X})\leq H(\epsilon)+\epsilon\log(|\mathcal{X}|-1) + H(Y)\\
\Rightarrow~~ &\log \binom{n}{d} \le-2\epsilon\log\epsilon + \epsilon\log\binom{n}{d} +T(1+\epsilon)\left(\frac{\rho d}{n}\log\left(\frac{n}{\rho d}\right)\right) \\
\Rightarrow~~ &(1-\epsilon)\log\binom{n}{d}+2\epsilon\log\epsilon \le (1+\epsilon)T\left(\frac{\rho d}{n}\log\left(\frac{n}{\rho d}\right)\right)\\
\Rightarrow~~ &\frac{(1-\epsilon)\log\binom{n}{d}+2\epsilon\log\epsilon}{(1+\epsilon)\frac{\rho d}{n}\log\left(\frac{n}{\rho d}\right)}\le T \\
\Rightarrow~~ & (1-5\epsilon)\frac{n}{\rho}\frac{\log\left(\frac{n}{d}\right)}{\log\left(\frac{n}{\rho d}\right)} \leq T.
\end{align*}
The last inequality follows from Stirling's approximation for $\binom{n}{d}$ (see Fact~\ref{fact:stirling}). 
Therefore,  for $\rho \in o(n/d)^{\beta}$, and sufficiently small $\epsilon >0$ at least $ T \geq (1-5\epsilon)\frac{n}{\rho}\frac{\log\left(\frac{n}{d}\right)}{\log\left(\frac{n}{\rho d}\right)} > \left( \frac{1-6\epsilon}{1-\beta}\right )\frac{n}{\rho}$ tests are needed.

%% file: row-construction.tex
\subsection{Proof of Theorem~\ref{thm:row-ub-random}: Randomized Construction of Test Matrices}
\label{sec:row:random}
\noindent
We now describe a randomized construction of a $T\times n$ test matrix $M$, where $T=c\left \lceil\frac{n}{\rho}\right \rceil $, where $c$ is a constant positive integer to be chosen later. For ease of exposition in this algorithm, it will help to assume that the probability of error $\epsilon$ scales as $n^{-\zeta}$ for some $\zeta >0$. Also recall that $d$ scales as $o(n^\alpha)$, and $\rho \in \Theta(n/d)^{\beta}$, for non-negative constants $\alpha$ and $\beta$ both less than $1$.
 
We pick $M$ by sampling uniformly from all $T \times n$ binary matrices with exactly $\rho$ items per test, and each item sampled exactly $c$ times.\footnote{The reason for sampling in this manner, rather than sampling rows uniformly at random from the set of all vectors with support exactly $\rho$, is because it makes analysis easier. 
}
We output the estimate vector $\widehat{X}$ from the test results using the same CoMa algorithm as in Section~\ref{subsubsec:coma}.

A fixed item is incorrectly marked defective when all the tests it participates in also correspond to tests in which at least one other defective item participates in, and hence is marked positive. 
While the choice of test matrix (uniform over $T \times n$ matrices with $c$ ones per column and $\rho$ ones per row) implies that any {\it individual} column is uniformly distributed among all length-$T$ columns of Hamming weight $c$. 
Since the total number of positive tests may be bounded from above by $cd$, the number of times each item is tested is exactly $c$, and any individual column of the test matrix is uniformly distributed among all length-$T$ columns of weight exactly $c$, therefore the probability that the item is incorrectly marked defective is at most $\frac{\dbinom{cd}{c}}{\dbinom{T}{c}}$.

\noindent
Taking a union bound over the $(n-d)$ nondefective items, we require 
\begin{equation}
(n-d)\frac{\dbinom{cd}{c}}{\dbinom{T}{c}}<\epsilon
\label{eq:rho:comp}
\end{equation} 
By Fact~\ref{fact:stirling} $\dbinom{cd}{c}<(ed)^{c}$ and ${\dbinom{T}{c}}>\left(\frac{T}{c}\right)^{c}$, and $T = cn/\rho$ by choice. Hence Equation~\eqref{eq:rho:comp} certainly occurs if 
\begin{equation}
\frac{n}{\epsilon} < \left (\frac{n}{\rho de} \right )^c. 	\label{eq:rho:comp2}
\end{equation}
Recalling that $\rho \in o(n/d)^{\beta}$ (hence $e\rho \ll (n/d)^{\beta}$),  Equation~\eqref{eq:rho:comp2} holds for $c > \log\left( \frac{n}{\epsilon}\right) / (1-\beta) \log \left(\frac{n}{d}\right)$.  Therefore,  recalling $d \in o(n^\beta)$ for some $0 < \beta <1$, and $\epsilon = n^{-\zeta}$, we have that Equation~\eqref{eq:rho:comp2} holds for sufficiently large $n$ if $c>\frac{1+\zeta}{(1-\alpha)(1-\beta)}$. Hence for sufficiently large $n$, $\left \lceil \frac{1+\zeta}{(1-\alpha)(1-\beta)} \right \rceil \left \lceil \frac{n}{\rho} \right \rceil$ tests suffice to guarantee a probability of error of at most $\epsilon = n^{-\zeta}$.

\begin{remark}
Note that the computational complexity of reconstruction of this algorithm, as in CoMa in Section~\ref{subsubsec:coma}, is $Tn$, which is in ${\cal O}(n^2/\rho)$.
\end{remark}

\begin{remark}\label{rem:row:ratio}
Note that this construction of a randomized test matrix that requires a number of tests that is larger by a factor of essentially $1/(1-\alpha)$ (neglecting dependencies on the probability of error) than the lower bound (in Theorem~\ref{thm:row-lb}) to reliably identify $d \in o(n^{\alpha})$ defective items. 
We note that this additional factor of $1/(1-\alpha)$ is similar to the state of affairs in classical group testing scenario (with no row/column constraints) (for instance see~\cite{ChanJSA14,ScarlettC16}).
\end{remark}

\begin{remark}\label{rem:universal}
Note that the design proposed in this section depends only very weakly on the specific value of $d$ -- the only place where the value of $d$ matters is in the constant pre-factor multiplying $n/\rho$ in the number of tests, and the requirement that $\rho \in o(n/d)$; the remainder of the design and reconstruction algorithm are independent of the specific value of $d$. 
Indeed, it can be directly verified that if one chooses $T \in \omega(n/\rho)$ (instead of $T \in \Theta(n/\rho)$ as in the design above), and one is guaranteed that $d \in o(n/\rho)$, then {\it regardless} of the specific value of $d$, the scheme above will {\it also} result in a design with a low probability of reconstruction error. 
This universality is reminiscent of {\it universal (fixed-length) source coding} -- as long as the input vector is ``sparse enough'', a given random code will be able to compress it in a manner compatible with low probability of reconstruction error.
\end{remark}

%% file: row-explicit.tex
\subsection{Proof of Theorem~\ref{thm:row-ub-explicit}: Explicit Construction of Test Matrices}
\label{sec:row:exp}
As in Section~\ref{sec:col:exp} we now provide an explicit design of test matrices for the scenario with $\rho$-sized tests. As in Section~\ref{sec:col:exp}, this explicit design only works for a restricted parameter range for $d$, when $d \in o(\sqrt{n})$. Specifically, let $d = \Theta(n^\alpha)$, $\alpha \in [0,1/2)$, $\rho \in \Theta(n^{(1-\alpha)\beta})$. 
The first tool in our explicit construction similar to the divide-and-conquer idea in Section~\ref{sec:col:exp}. We divide the $[n]$ items into at least $\frac{n\epsilon}{d^2}$ blocks, where each block contains at most $\rho$ items. 
Specifically: 
\begin{enumerate}
	\item 
	When $\rho < \frac{n\epsilon}{d^2}$, we divide $[n]$ into $\left \lceil \frac{n}{\rho} \right \rceil > \frac{d^2}{\epsilon}$ blocks, each of size at most $n' = \rho$.
	\item 
	When $\rho \geq \frac{n\epsilon}{d^2}$, we divide $[n]$ into $\left \lceil \frac{d^2}{\epsilon} \right \rceil$ blocks, each of size at most $n' = \left \lceil \frac{n\epsilon}{d^2} \right \rceil \leq \rho$.
\end{enumerate}
Since in both regimes, by design, there are at least $\frac{d^2}{\epsilon}$ blocks, therefore as in Section~\ref{sec:col:exp}, with probability at least $1-\epsilon$, each group contains at most one defective item.

Within each block, our test-design is a ``non-adaptive binary search''. Specifically, for each $i$th block of size $n'$, there is a $\lceil \log(n'+1)\rceil \times n'$ ``sub-test matrix'' $M^{(i)}$ that non-adaptively tests only the corresponding $n'$ items. The $j$th column of $M^{(i)}$ comprises of the (length-$\lceil \log(n'+1)\rceil$) binary representation of the ingteger $j$. The decoding algorithm, on observing test outcomes corresponding to block $i$, declares no defectives present if all test outcomes are negative. On the other hand, if some test outcomes are positive, then viewing positive test outcomes as $1$s and negatives test outcomes as $0$s, the length-$\lceil \log(n'+1)\rceil$ test outcome vector viewed as the binary representation of integer $j$ precisely identifies the unique defective item in block $i$. Hence, conditioned on each block containing at most one defective item (which happens with probability at least $1-\epsilon$), the above algorithm always outputs the correct answer.

The number of tests and computational complexity of reconstruction of this design are as follows:
\begin{enumerate}
	\item 
	When $\rho < \frac{n\epsilon}{d^2}$, the number of tests and computational complexity of reconstruction are both at most $\left \lceil \frac{n}{\rho} \right \rceil \lceil \log(\rho+1)\rceil $.
	\item 
	When $\rho \geq \frac{n\epsilon}{d^2}$, the number of tests and computational complexity of reconstruction are both at most $\left \lceil \frac{d^2}{\epsilon} \right \rceil \lceil \log \left (\frac{n\epsilon}{d^2}+1 \right )\rceil $.
\end{enumerate}

\begin{remark}
Note that in the first regime, the number of tests required exceed the information-theoretic lower bound derived in Theorem~\ref{thm:row-lb} by only a factor of about $\log(\rho)$. 
On the other hand, in the second regime the number of tests required is larger than the {\it unconstrained} information-theoretic lower bound $\Theta(d\log(n))$ by a factor of $\Theta(d)$, which may be as large as ${\cal O}(\sqrt{n})$ (since in our design $d$ is restricted to be in ${\cal O}(\sqrt{n})$). 
As in Section~\ref{sec:col:exp}, test designs that reduce these extra factors (and obviate the need to restrict $d$ to be in ${\cal O}(\sqrt{n})$) is an interesting open question.
\end{remark}

%% file: noise.tex
\section{Impact of Noisy Tests}
\label{sec:noise}
We now consider the impact of noise in test outcomes on the performance on group-testing algorithms. 
While multiple noise models (for instance, erasures~\cite{cohen2016secure}, bit-flips~\cite{atia2012boolean,scarlett2017limits,ChanJSA14}, dilution noise~\cite{atia2012boolean}) have been considered in the literature, for the sake of concreteness, we focus on perhaps the most commonly considered model in the literature, bit-flip noise, wherein test outcomes are passed through a binary-symmetric channel with crossover probability $0<\sigma < 1/2$.
\begin{remark}
At first sight, it might seem surprising that the model with $\rho$-size constraints on tests allows for designs that are robust to noise, but the model with $\gamma$-divisibility constraints on items does not. 
The underlying reason for this asymmetry is that if an item only participates in relatively few tests (as in the $\gamma$-divisible model), then with non-trivial probability evidence of its status (defective or not) can be erased. 
On the other hand, even if tests are highly constrained in size (as in the $\rho$-sized tests model), relatively simple ideas like repetition coding (as described in Section~\ref{thm:row-noise} allow for each individual test to be made highly reliable -- this option is unavailable in the $\gamma$-divisible setting.
\end{remark}

\subsection{Proof of Theorem~\ref{thm:column-noise}: $\gamma$-Divisible Items}\label{subsec:column-noise}

\newcommand{\hX}{{\hat{X}}}
\newcommand{\hY}{{\hat{Y}}}

\newcommand{\hXi}{{\hat{X}({i}^c)}}

We first consider the noisy setting, where each test can be incorrect with probability $0<\sigma<1/2$, for $\gamma$-divisible tests. Recall that, for ease of analysis, solely in this section our probability distribution over the set ${\cal D}$ of defectives is that each item is defective with probability $d/n$ in an i.i.d. manner.

Given the observed length-$T$ vector of noisy test outcomes $\hat{Y}$, say the decoder outputs the length-$n$ vector $\hat{X}$. Let $\hXi$ denote the length-$n$ vector that equals $\hat{X}$ in each coordinate except the $i$th, in which location the value $(\hXi)_i$ equals $1- (\hat{X})_i$, i.e., corresponds to $i$th bit of $\hat{X}_i$ being flipped. 
Roughly speaking, we will now show the set of these $\hXi$s is ``relatively easily confusable'' with $\hX$. More precisely, our strategy will be to show that for {\it any} test matrix with at most $\gamma$ ones per column, any $\hY$, and any $\hX$ (corresponding to a decoding strategy mapping $\hY$ to $\hX$), the ratio
\begin{equation}
	r = \frac{\sum_{i=1}^n \Pr(X = \hXi|\hY)}{\Pr(X = \hX|\hY)} \label{eq:pe-low}
\end{equation}
is ``non-trivial". 
That is, given its observation $\hY$, regardless of how the decoder picks his estimate $\hX$, the probability that the true $X$ corresponded to one of $\hXi$ is at least a factor $r$ of the probability that $X$ equaled $\hX$. 
This would imply a lower bound on the probability of decoding error as 
\begin{equation}
\Pr(error)=	\Pr(\hX \neq X) \geq \frac{r}{r+1}.\label{eq:pe-lbr}
\end{equation}
Hence as long as one can bound $r$ by a quantity asymptotically bounded away from zero, one can bound the probability of error away from zero. To do so, we individually bound each of the $n$ terms in Equation~\eqref{eq:pe-low}.

By Bayes' rule, for any $i$, 
\begin{eqnarray}
\frac{\Pr(X = \hXi|\hY)}{\Pr(X = \hX|\hY)} & = &	 \frac{\Pr(\hY|X = \hXi)}{\Pr(\hY|X = \hX)}\frac{\Pr(X = \hXi)}{\Pr(X = \hX)} \label{eq:bayes}
\end{eqnarray}
We now bound from below the two factors $\frac{\Pr(\hY|X = \hXi)}{\Pr(\hY|X = \hX)}$ and $\frac{\Pr(X = \hXi)}{\Pr(X = \hX)}$ in Equation~\eqref{eq:bayes}

Noting that $\hX$ and $\hXi$ differ in just the $i$th bit, the ratio $\frac{\Pr(X = \hXi)}{\Pr(X = \hX)}$ equals either $\frac{d/n}{1-d/n}$ (if $X_i = 0$) or $\frac{1-d/n}{d/n}$ (if $X_i = 1$), and hence we have the lower bound $d/(n-d) > d/n$.
\begin{equation}
	\frac{\Pr(X = \hXi)}{\Pr(X = \hX)} \geq \frac{d}{n-d} > \frac{d}{n}. \label{eq:bays-lb1}
\end{equation}

We next consider the term $\frac{\Pr(\hY|X = \hXi)}{\Pr(\hY|X = \hX)}$. 
Let $Y_M(\hXi)$ denote the {\it noiseless} length-$T$ test outcome vector corresponding to test matrix $M$ and input $\hXi$, and $Y_M(\hX)$ denote the noiseless length-$T$ test outcome vector corresponding to test matrix $M$ and input $\hX$. 
Since, for a fixed $M$, $X \leftrightarrow Y \leftrightarrow \hY \leftrightarrow \hX$ form a Markov chain, the ratio $\frac{\Pr(\hY|X = \hXi)}{\Pr(\hY|X = \hX)}$ equals $\frac{\Pr(\hY|Y_M(\hXi))}{\Pr(\hY|Y_M(\hX))}$.  
Since the test matrix is constrained to have at most $\gamma$ ones per column and $\hX$ and $\hXi$ differ in just a single bit, therefore $Y_M(\hX)$ and $Y_M(\hXi)$ differ in at most $\gamma$ locations. 
Hence we have the lower bound
\begin{equation}
\frac{\Pr(\hY|X = \hXi)}{\Pr(\hY|X = \hX)} =	\frac{\Pr(\hY|Y_M(\hXi))}{\Pr(\hY|Y_M(\hX))} \geq \left (\frac{\sigma}{1-\sigma} \right )^\gamma. \label{eq:bays-lb2}
\end{equation}

Substituting Equations~\eqref{eq:bays-lb1} and~\eqref{eq:bays-lb2} into Equation~\eqref{eq:bayes}, and noting that there are $n$ terms in Equation~\eqref{eq:pe-low}, allows us to bound $r$ from below as $d\left (\frac{\sigma}{1-\sigma} \right )^\gamma$. Hence if $\log(d)/\gamma \in \Omega(1)$, the probability of error is bounded away from zero.

\subsection{Proof of Theorem~\ref{thm:row-noise}: $\rho$-Sized Tests}
Finally, we consider the setting where each test can be incorrect with probability $0<\sigma<1/2$, for $\rho$-sized tests. 

The idea here is quite straightforward -- we use essentially the same design as in Section~\ref{sec:row:random}, but each test is repeated $k$ times, for a design parameter $k$ specified below. 
For each set of $k$ repeated tests, the reconstruction algorithm takes the majority outcome to represent the outcome of the single test in the noiseless model, and thence uses the reconstruction algorithm in Section~\ref{sec:row:random}. 
Since each test is misreported with probability $\sigma$, the expected number of failures over $k$ repetitions is $k\sigma$. 
Then by standard Chernoff bounds, the probability that the number of failures is at least $k/2$ is at most $\exp \left (-\frac{k(1/2-\sigma)^2}{2\times \frac{1}{2}} \right) = \exp \left (-{k(1/2-\sigma)^2} \right)$. 
Hence, if $k=\left \lceil \frac{(1+\zeta)\ln(n)}{(1/2-\sigma)^2}\right \rceil$, the probability that any single majority outcome is incorrect is at most $\lceil n^{-1-\zeta} \rceil$. 
Since the number of tests in the design in Section~\ref{sec:row:random} is $\Theta\left (\frac{n}{\rho}\right ) \in o(n)$, therefore the probability that any of the $o(n)$ majority test outcomes is less than $n^{-\zeta}$. 
Hence the overall probability of error is at most $2n^{-\zeta}$ ($n^{-\zeta}$ from the probability of an incorrect majority outcome, and $n^{-\zeta}$ from the probability of reconstruction error in the algorithm in Section~\ref{sec:row:random}), the overall number of tests is at most $k$ times the number of tests in the design in Section~\ref{sec:row:random}, i.e. $\left \lceil \frac{1+\zeta}{(1-\alpha)(1-\beta)} \right \rceil\left \lceil \frac{n}{\rho}\right \rceil \left \lceil \frac{(1+\zeta)\ln(n)}{(1/2-\sigma)^2}\right \rceil$, and the overall reconstruction complexity is at most ${\cal O}\left (\frac{n^2\log(n)}{\rho}\right )$.
\begin{remark}\label{rem:nolog}
While the repetition coding scheme presented above has the advantage of relative simplicity of presentation and analysis, it does require a number of tests that is larger than the noiseless scenario by a multiplicative factor of $\log(n)$. 
In principle, however, it is conceivable that the Noisy CoMa algorithm in~\cite{ChanJSA14} will also work in the $\rho$-sized noisy test setting. 
To avoid the intricate calculations required to validate this suggested code-design, in this journeyman work on sparse group testing we leave open this possibility for future work.
\end{remark}

%% file: futurework.tex
\section{Conclusion}\label{sec:futwork}
In this work we consider the impact of constraints on non-adaptive group-testing the number of times items can be tested, or the size of tests. 
In both settings we show that even mild constraints can result in a dramatic blowup (compared to the unconstrained setting) in the number of tests required, and provide algorithms with computationally efficient reconstruction algorithms that (nearly) match the performance (in terms of the number of tests required for reliable reconstruction) of the lower bounds we prove. 
We also consider the impact of noisy test outcomes.\\

%% file: acknowledgements.tex
\section*{Acknowledgements}
This research was initiated at the ICERM Workshop on Algorithmic Coding Theory, 2016. We would like to thank ICERM and the organizers for their hospitality. We would also like to thank Abhinav Ganesan and Gu Wenyuan for fruitful discussions, and the anonymous referees and Associate Editor for insightful comments on an earlier draft of this work. S. J was supported in part by a Google grant and GRF grant 14304418; V. G., E. G., and S. Z. were supported in part by NSF CCF-1649515. E. G. was also supported in part by a grant from the Purdue Research Foundation.